\documentclass[fleqn,10pt]{wlscirep}
\usepackage[utf8]{inputenc}
\usepackage[T1]{fontenc}
\usepackage{bm}
\usepackage{here}
\graphicspath{{./pics/}}
\newcommand*{\sigmaf}{\sigma^\mathrm{(fric)}}
\newcommand*{\Pext}{P_\mathrm{ext}}
\newcommand*{\vc}{v_\mathrm{c}}
\newcommand*{\muS}{\mu_\mathrm{S}}
\newcommand*{\muK}{\mu_\mathrm{K}}
\newcommand*{\muM}{\mu_\mathrm{M}}
\newcommand*{\Ac}{A_\mathrm{c}}
\newcommand*{\alphaA}{\alpha_\mathrm{\,A}}
\newcommand*{\alphaB}{\alpha_\mathrm{\,B}}
\newcommand*{\etat}{\eta_\mathrm{\,t}}
\newcommand*{\Vrod}{V_\mathrm{rod}}

\newcommand{\tcgreen}[1]{\textcolor{black}{#1}}
\newcommand{\tcmagenta}[1]{\textcolor{black}{#1}}
\title{Static Friction Coefficient Depends on the External Pressure and Block Shape due to Precursor Slip}

\author[1,*]{Wataru Iwashita}
\author[2]{Hiroshi Matsukawa}
\author[1]{Michio Otsuki}
\affil[1]{Osaka University, Department of Mechanical Science and Bioengineering, Toyonaka, 560-8531, Japan}
\affil[2]{Aoyama Gakuin University, Department of Physical Sciences, Sagamihara, 252-5258, Japan}

\affil[*]{w\_iwashita@fm.me.es.osaka-u.ac.jp}

\begin{abstract}
Amontons' law states that the maximum static friction force on a solid object is proportional to the loading force and is independent of the apparent contact area. 
This law indicates that the static friction coefficient does not depend on the external pressure or object shape. 
Here, we numerically investigate the sliding motion of a 3D viscoelastic block on a rigid substrate using the finite element method (FEM). 
The macroscopic static friction coefficient decreases with an increase in the external pressure, length, or width of the object, which contradicts Amontons' law. 
Precursor slip occurs in the 2D interface between the block and substrate before bulk sliding. 
The decrease in the macroscopic static friction coefficient is scaled by the critical area of the precursor slip before bulk sliding. 
A theoretical analysis of the simplified models reveals that bulk sliding results from the instability of the quasi-static precursor slip caused by velocity-weakening local friction. 
We also show that the critical slip area determines the macroscopic static friction coefficient, which explains the results of the FEM simulation. 
\end{abstract}

\begin{document}

\flushbottom
\maketitle

\thispagestyle{empty}

%%%%%%%%%%%%%%%%%%%%%%%%
%%%%% Introduction %%%%%
%%%%%%%%%%%%%%%%%%%%%%%%
\section*{Introduction}
%%%%%%%%%%%%%%%%%%%%%%%%

A friction force prevents the relative sliding motion between two objects in contact. 
Friction plays a crucial role in various situations, such as the contact surface between the ground and the sole of a shoe, brakes and bearings in machines, and tectonic plates that cause earthquakes. 
Many studies on friction have been conducted, but the elucidation of the fundamental mechanism of friction is essential for science and technology~\cite{Bowden1950, Persson2000, Popov2017_text, Rabinowicz1995, Dowson1998, Bhushan2013, Baumberger2006}.

Amontons' law states that the maximum static friction force on a solid object is independent of the apparent contact area and proportional to the load~\cite{Bowden1950, Persson2000, Popov2017_text, Rabinowicz1995, Dowson1998, Bhushan2013, Baumberger2006}. 
This law has been taught in high school physics textbooks and is believed to hold true for diverse systems.
When the friction force obeys Amontons' law, the friction coefficient, which is the ratio of the friction force to the loading force, does not depend on the pressure, size, or object shape. 
On a rough frictional interface with \tcgreen{numerous} asperities, only a tiny fraction of the surfaces \tcgreen{forms} junctions, the so-called real contact points.
Amontons’ law is explained by the proportionality of the total area of real contact points to the loading force~\cite{Bowden1950, Persson2000, Popov2017_text, Rabinowicz1995, Dowson1998, Bhushan2013, Baumberger2006, Archard1957, Dieterich1996}.

The above explanation for the origin of Amontons' law implicitly assumes uniformity of the stress field. 
Therefore, Amontons' law is not expected to hold if a macroscopic deformation exists. 
In fact, recent numerical studies have reported the breakdown of Amontons' law in macroscopic viscoelastic objects~\cite{Otsuki2013,Ozaki2014}, revealing that it is related to local quasi-static precursor slips before the onset of bulk sliding owing to non-uniform deformation~\cite{Otsuki2013, Katano2014, Bouissou1998, Rubinstein2004, Rubinstein2007, Ben-David2010, Ben-David2011, Malthe-Sorenssen2021, Braun2009, Maegawa2010, Scheibert2010, Amundsen2012, Tromborg2011, Tromborg2014, Radiguet2013, Kammer2015, Ozaki2014, Taloni2015, deGeus2019}.
The relationship between precursor slips and the breakdown of Amontons' law \tcgreen{has been} confirmed \tcgreen{previously} in an experiment with an acrylic glass block~\cite{Katano2014}. 
However, previous studies have only investigated systems with a 1D frictional interface.
Friction usually occurs in 2D interfaces of 3D objects.
However, it is not clear whether the results in previous studies apply to more realistic 3D systems.

In this study, we numerically investigate the sliding motion of a 3D viscoelastic object on a rigid substrate using the finite element method (FEM). 
The macroscopic static friction coefficient decreases with an increase in the pressure or size of the object. 
The precursor slip propagates in a 2D frictional interface. 
Bulk sliding occurs when the area of the precursor slip reaches a critical value, which determines the macroscopic static friction coefficient. 
An analysis of the simplified models reveals that the instability of the precursor slip leads to bulk sliding.

%%%%%%%%%%%%%%%%%%%
%%%%% Results %%%%%
%%%%%%%%%%%%%%%%%%%
\section*{Results}
%%%%%%%%%%%%%%%%%%%

%%%%%%%%%%%%%%%%%%
%%%%% 3D FEM %%%%%
%%%%%%%%%%%%%%%%%%
\subsection*{3D FEM simulation}

\begin{figure}[H]
    \centering
    \includegraphics[scale=1.]{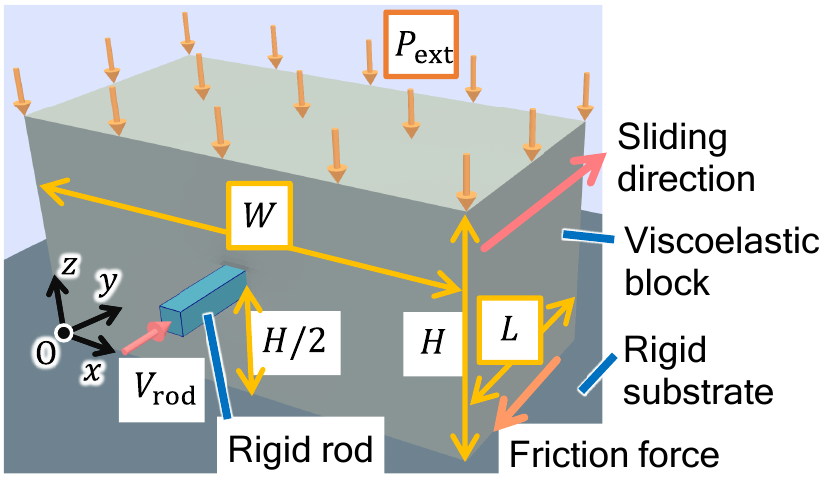}
    \caption{Schematic of a 3D viscoelastic block on a fixed rigid substrate. }
    \label{fig:1_schem_block}
\end{figure}

We numerically investigate a viscoelastic block on a rigid substrate with width $W$, length $L$, and height $H$ along the $x$-, $y$-, and $z$-axes, respectively, as shown in Fig.~\ref{fig:1_schem_block} (see Methods for details). 
The area of the frictional interface is denoted by $A_0 = L W$.
The density, Young's modulus, and Poisson's ratio of the block are denoted \tcgreen{by} $\rho$, $E$, and $\nu$, respectively.
The dissipation in the block is characterized by two viscosity coefficients: $\eta_1$ and $\eta_2$. 
We assume that Amontons' law holds locally at the interface between the block and the rigid substrate ($z=0$), and the magnitude of the local frictional stress, $\sigmaf(x, y)$ in the interface is locally determined as
\begin{equation}
    \sigmaf(x, y) = \mu(v(x, y)) p(x, y), 
    \label{eq:sigma_fric}
\end{equation}
where $p(x, y)$ is the bottom pressure\tcgreen{,} and $\mu(v)$ is the friction coefficient, which depends on the magnitude of the local slip velocity $v(x, y)$ when $v(x,y)\neq0$~~\cite{ccm2006}. 
\tcmagenta{
Here, $\mu(v)$ is characterized by the characteristic velocity of velocity-weakening friction $\vc$ and the local static and kinetic friction coefficients denoted by $\muS$ and $\muK$ (see Methods). 
The rigid rod quasi-statically pushes the center of the side surface along the $y$ direction. 
}
\tcgreen{The} macroscopic friction force $F_\mathrm{T}$ is measured as the force on the rigid rod in the $y$ direction. 
The loading force applied to the top of the block is given by $F_\mathrm{N} = \Pext A_0$ \tcmagenta{with the external pressure to the top surface $\Pext$}. 

%%%%%
\begin{figure}[H]
    \centering
    \includegraphics[scale=1.]{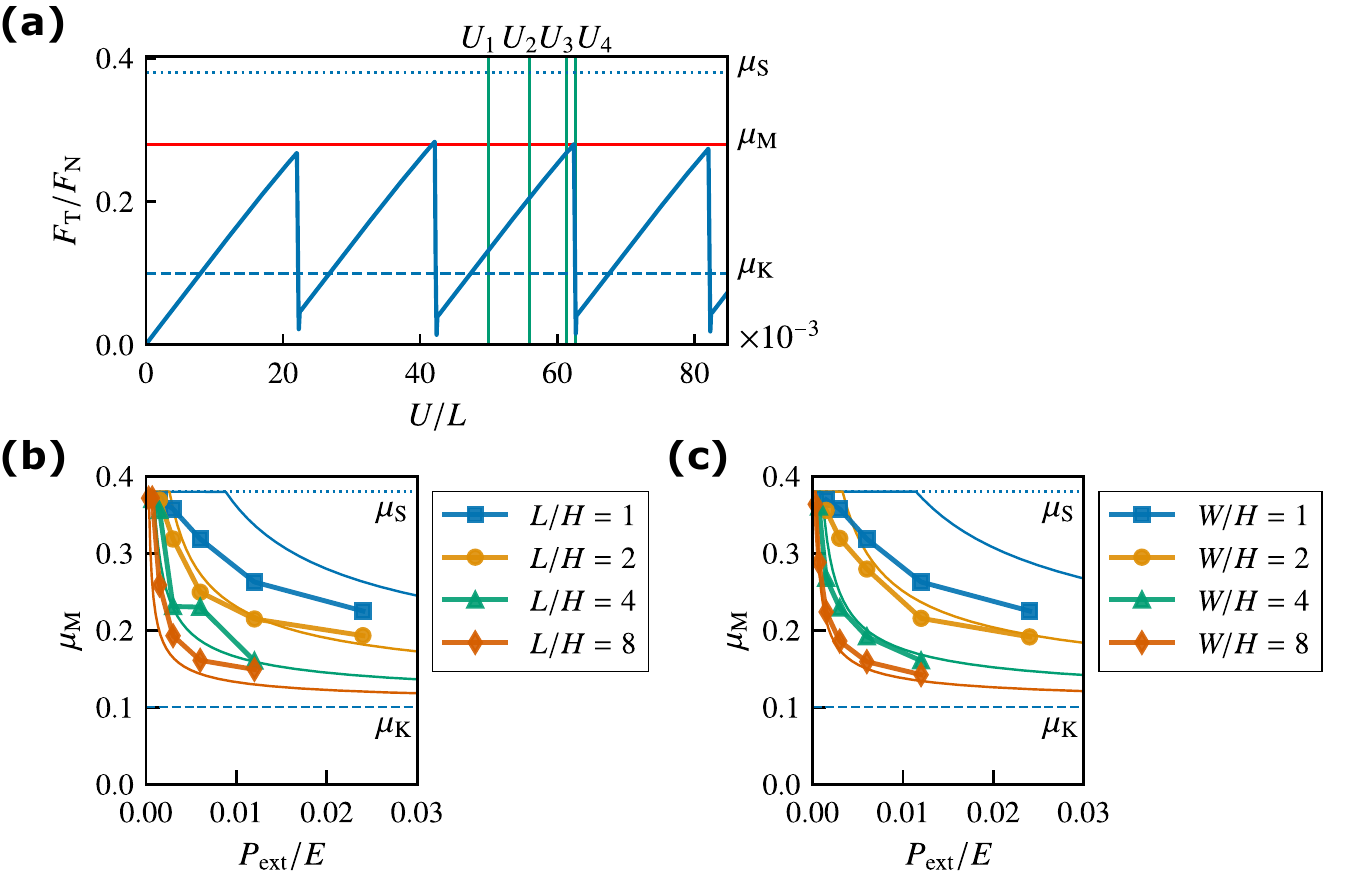}
    \caption{(a)~Ratio $F_\mathrm{T}/F_\mathrm{N}$ against the displacement of the rigid rod $U$ for $L/H = 1$, $W/H = 2$, and $\Pext/E = 0.006$. 
    The red horizontal line represents the macroscopic static friction coefficient $\muM$. 
    (b)~Macroscopic static friction coefficient $\muM$ against pressure $\Pext$ for various $L/H$ values with $W/H = 1$.
    The thin solid lines represent the analytical results with $\alphaA = 0.2$ given by equations~\eqref{eq:lc_p_1dl_ver1} and \eqref{eq:mu_lc_1dl}.
    (c)~Macroscopic static friction coefficient $\muM$ against $\Pext$ for various $W/H$ values with $L/H = 1$.
    The thin solid lines represent the analytical results with $\alphaB = 0.2$ given by equations~\eqref{eq:mu_lc_1dl} and \eqref{eq:lc_p_1dw_ver1}.
    The dotted and dashed lines represent $\muS$ and $\muK$, respectively.
    }
    \label{fig:2_friction_force}
\end{figure}
%%%%%
%%%%%
\begin{figure}[H]
    \centering
    \includegraphics[scale=1.]{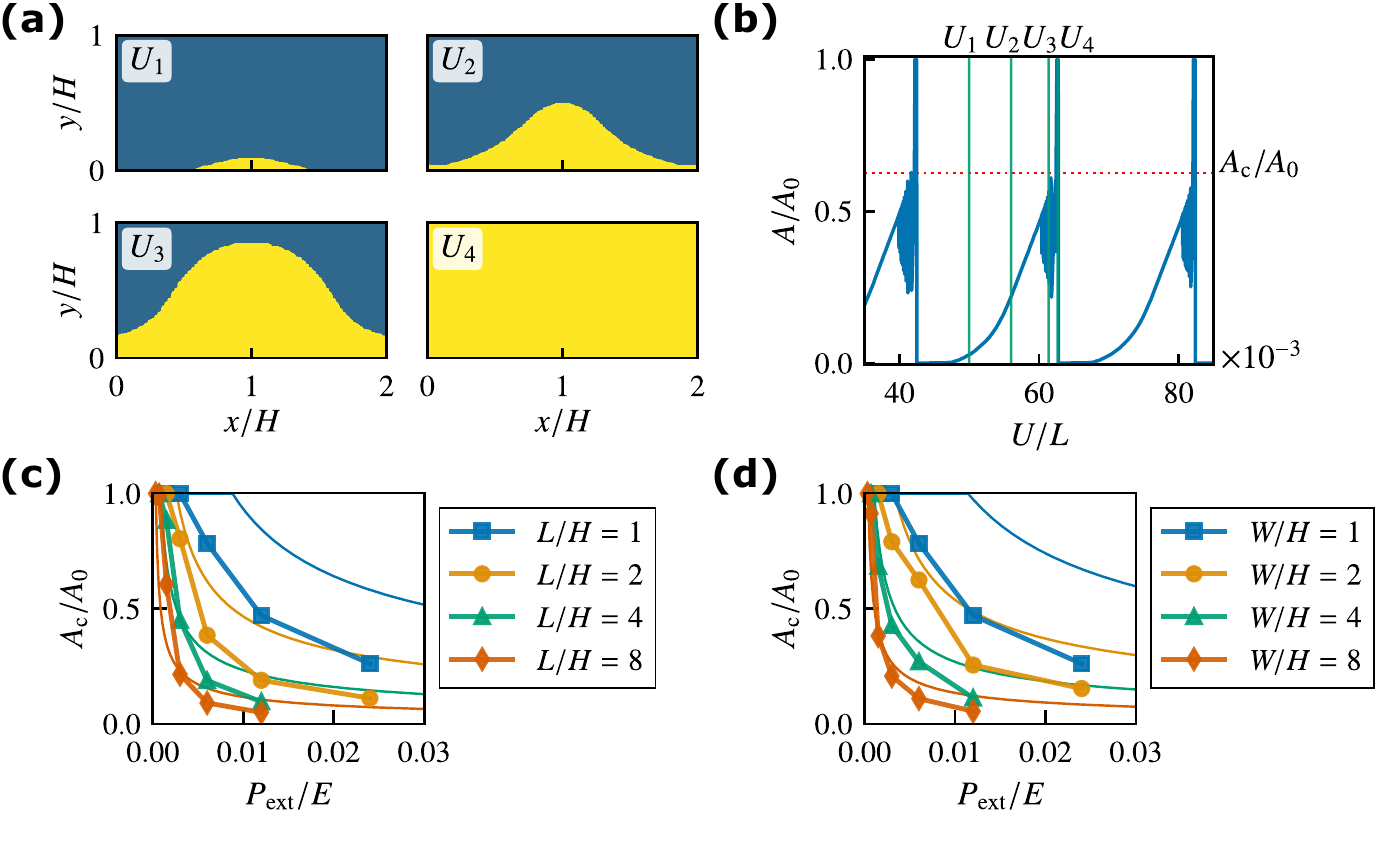}
    \caption{(a)~Spatial distribution of the slip region in the frictional interface at $U=U_1, U_2, U_3$, and $U_4$ for $L/H = 1$, $W/H = 2$, and $\Pext/E = 0.006$. 
    The yellow area represents the slip region. The rigid rod is pushing the block at $(x/H,y/H)=(1,0)$. 
    (b)~Normalized precursor slip area $A/A_0$ against displacement $U$. 
    The dotted line represents the normalized critical area $\Ac/A_0$. 
    (c)~Normalized critical area $\Ac/A_0$ against pressure $\Pext$ for various $L/H$ values with $W/H = 1$. 
    The thin solid lines represent the analytical results with $\alphaA = 0.2$ given by equation~\eqref{eq:lc_p_1dl_ver1}.
    (d)~Normalized critical area $\Ac/A_0$ against $\Pext$ for various $W/H$ values with $L/H = 1$. 
    The thin solid lines represent the analytical results with $\alphaB = 0.2$ given by equation~\eqref{eq:lc_p_1dw_ver1}.
    }
    \label{fig:3_precursor_slip}
\end{figure}
%%%%%

The ratio $F_\mathrm{T}/F_\mathrm{N}$ is plotted against the displacement of the rigid rod $U$ for $L/H = 1$, $W/H = 2$, and $\Pext/E = 0.006$ in Fig.~\ref{fig:2_friction_force}~(a).
First, $F_\mathrm{T}/F_\mathrm{N}$ increases linearly with $U$.
After obtaining a maximum value lower than $\muS$, $F_\mathrm{T}/F_\mathrm{N}$ rapidly decreases to a value close to $\muK$. 
This rapid drop is associated with bulk sliding. 
The significant drop after the linear increase periodically repeats \tcgreen{itself}. 
This periodic behavior corresponds to the stick-slip motion of the object.
The maximum value of $F_\mathrm{T}/F_\mathrm{N}$ represents the macroscopic static friction coefficient, $\muM$. 
Figures~\ref{fig:2_friction_force}~(b) and (c) display the macroscopic static friction coefficient $\muM$ against pressure $\Pext$ for various $L/H$ and $W/H$ values, respectively. 
The magnitude of $\muM$ decreases with increasing $\Pext$, which is qualitatively consistent with the results for a system with a 1D friction interface in ref.~\citenum{Otsuki2013}. 
The previous study reported the size dependence of $\muM$ while maintaining the aspect ratio $L/H=2$~\cite{Otsuki2013}, whereas Figs.~\ref{fig:2_friction_force}~(b) and (c) demonstrate that the friction coefficient $\muM$ also decreases with increasing aspect ratios $L/H$ and $W/H$.
These results indicate that Amontons' law breaks down in systems with 2D interfaces.

Figure~\ref{fig:3_precursor_slip}~(a) shows the spatial distribution of the slip region with nonzero slip velocity in the frictional interface at $z=0$ for $U=U_1, U_2, U_3$, and $U_4$ shown in Fig.~\ref{fig:2_friction_force}~(a).
Here, we choose $U_1/L=50\times10^{-3}$, $U_2/L=56\times10^{-3}$, $U_3/L=61.38\times10^{-3}$, and $U_4/L=62.71\times10^{-3}$, \tcgreen{which corresponds to the stationary stick-slip region}. 
See Methods for the definition of the slip region. 
In Fig.~\ref{fig:3_precursor_slip}~(a), the local precursor slip starts from the region under the rigid rod for $U=U_1$.
As $U$ increases ($U_2$ and $U_3$), the region expands \tcgreen{gradually}. 
After $U=U_3$, the entire area slips with $v>\vc$, resulting in bulk sliding.
Note that the slip occurs almost along the $y$ direction.
Figure~\ref{fig:3_precursor_slip}~(b) shows the area of precursor slip $A$ normalized by the area of frictional interface $A_0$ against displacement $U$. 
First, the area of the precursor slip increases \tcgreen{gradually} with displacement $U$.
When \tcgreen{the} area $A$ reaches the critical area $\Ac$ just before bulk sliding (dotted line), the propagation speed of the area suddenly increases. 
Owing to rapid propagation, $A$ reaches $A_0$ and then returns to $0$.
We demonstrate the normalized critical area $\Ac/A_0$ against pressure $\Pext$ in Figs.~\ref{fig:3_precursor_slip}~(c) and (d) for various $L/H$ values with $W/H=1$ and for various $W/H$ values with $L/H=1$, respectively.
The normalized critical area $\Ac/A_0$ decreases as $\Pext$, $L/H$, or $W/H$ increases.
This decrease is similar to that of $\muM$ in Figs.~\ref{fig:2_friction_force}~(b) and (c), respectively.

\begin{figure}[H]
    \centering
    \includegraphics[scale=1.]{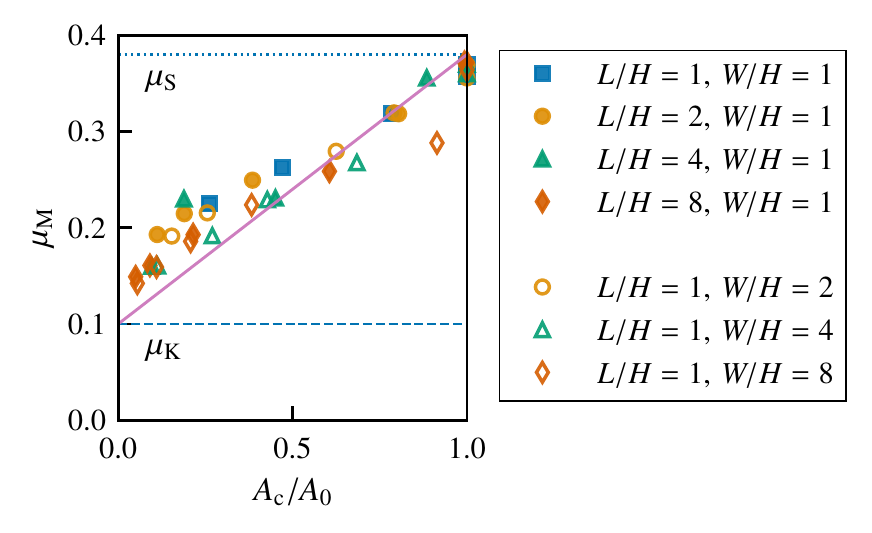}
    \caption{Macroscopic static friction coefficient $\muM$ against the normalized critical area $\Ac/A_0$ for various $L/H$ and $W/H$ values.
    The solid line represents the analytical result given by equation~\eqref{eq:mu_lc_1dl}.
    The dotted and dashed lines represent $\muS$ and $\muK$, respectively. }
    \label{fig:4_muM_Ac}
\end{figure}

In Fig.~\ref{fig:4_muM_Ac}, we present the macroscopic friction coefficient $\muM$ against the normalized critical area $\Ac/A_0$ for various $L/H$ and $W/H$ values.
The macroscopic friction coefficient $\muM$ for different $L/H$ and $W/H$ values approximately collapses onto a master curve, which indicates a linear increase in $\muM$ with $\Ac/A_0$. 
The minimum value close to $\Ac/A_0 = 0$ is almost equal to $\muK$, whereas the maximum value at $\Ac/A_0=1$ is equal to $\muS$.

%%%%%
%%%%%
%%%%%
\begin{figure}[H]
    \centering
    \includegraphics[scale=1.]{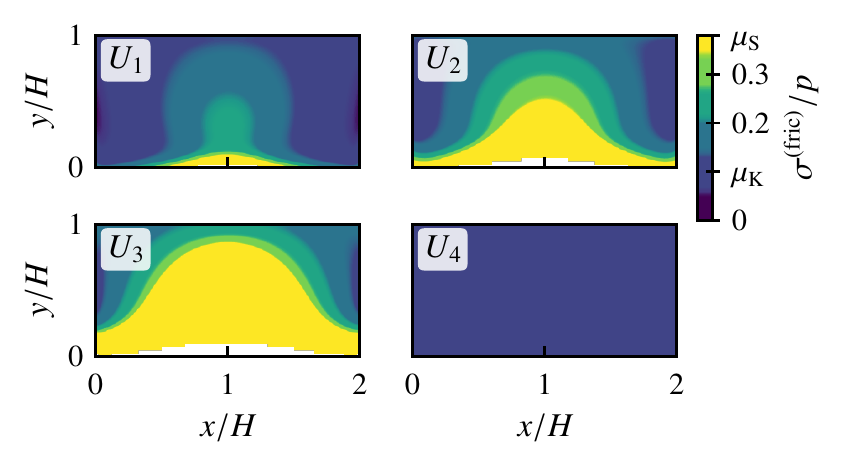}
    \caption{Spatial distribution of the ratio of frictional stress to bottom pressure $\sigmaf/p$ in the frictional interface for $L/H = 1$, $W/H = 2$, and $\Pext/E = 0.006$ at $U=U_1, U_2, U_3$, and $U_4$.
    The rigid rod is pushing the block at $(x/H,y/H)=(1,0)$. 
    The white area represents the region with $p=0$ due to the lift of the bottom. }
    \label{fig:5_sfric_p}
\end{figure}
%%%%%

Figure~\ref{fig:5_sfric_p} shows the spatial distribution of the ratio $\sigmaf/p$ in the frictional interface for $L/H = 1$, $W/H = 2$, and $\Pext/E = 0.006$ at $U= U_1, U_2, U_3$, and $U_4$.
It should be noted that the direction of the frictional stress is almost opposite to the driving direction, that is, the $y$ direction. 
\tcmagenta{
In the no-slip region, the local static friction can take any value for $0<\sigmaf/p<\muS$.
}
Before the onset of precursor slip, that is, just after bulk sliding, $\sigmaf/p$ takes a value almost equal to $\muK$, the local kinetic friction coefficient, in the entire interface, as explained below. 
At $U=U_1$, $\sigmaf/p$ reaches the local static friction coefficient, $\muS$\tcgreen{,} near the rigid rod at $(x/H,y/H)=(1,0)$.
As the displacement $U$ increases to $U_2$ and $U_3$, the area with $\sigmaf/p \simeq \muS$ \tcgreen{gradually} increases. 
The region of $\sigmaf/p \simeq \muS$ coincides with the local precursor slip region in Fig.~\ref{fig:3_precursor_slip}~(a). 
Except for the slip region, $\sigmaf/p$ remains \tcgreen{approximately at} $\muK$. 
Immediately after $U_3$, bulk sliding with $v>\vc$ occurs, and the fast slip leads to $\sigmaf/p = \muK$ at $U_4$. 
Bulk sliding rapidly \tcgreen{decelerates,} and the slip velocity $v$ decreases to $0$\tcgreen{,} when $\sigmaf/p$ \tcgreen{increases} to $\muS$ in the frictional interface.
\tcmagenta{
However, the internal deformation is not able to follow the rapid change, and the ratio of static frictional stress to bottom pressure finally returns to $\sigmaf/p \simeq \muK$ after bulk sliding. 
}
Consequently, $\sigmaf/p$ is almost equal to $\muK$ after bulk sliding. 
The macroscopic static friction coefficient $\muM$ is approximately expressed by the average of $\sigmaf/p$ over the entire frictional interface at $U_3$ immediately before bulk sliding.
This result explains the dependence of $\muM$ on $\Ac/A_0$ shown in Fig.~\ref{fig:4_muM_Ac}, where $\muM$ approaches $\muS$ for $\Ac/A_0=1$.

%%%%%%%%%%%%%%%%%%%%%%%%%%%%%%%%%%%%%
%%%%% Simplified model analysis %%%%%
%%%%%%%%%%%%%%%%%%%%%%%%%%%%%%%%%%%%%
\subsection*{Analysis based on simplified models}

To theoretically analyze the numerical results, we employ two simplified models, which explain the dependence of $\muM$ on $L/H$ and $W/H$ (see Supplementary Note online for details). 

%%%%%
\begin{figure}[tb]
    \centering
    \includegraphics[scale=1.]{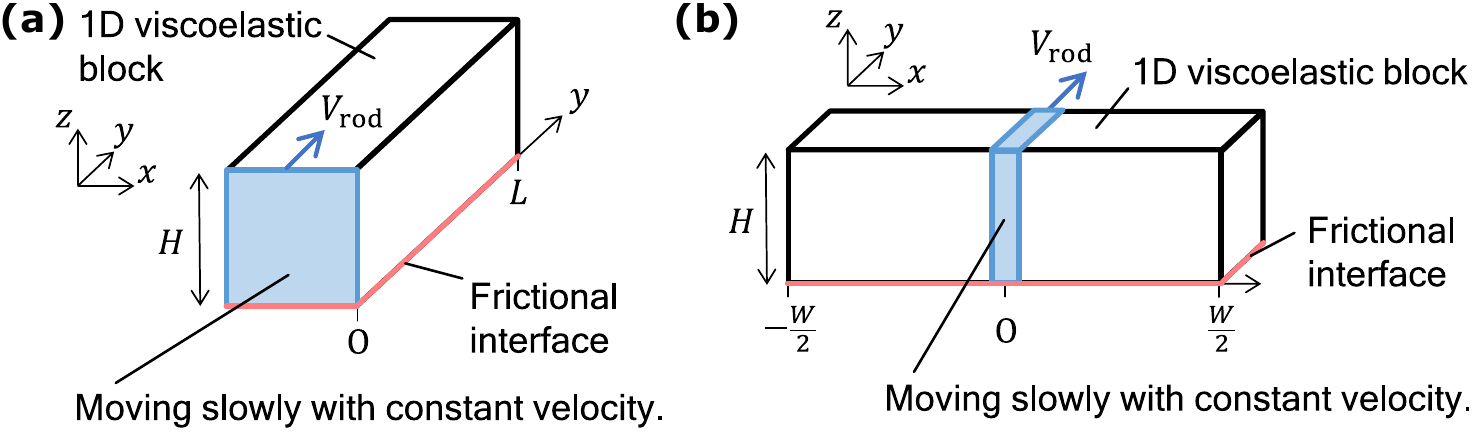}
    \caption{Schematics of simplified models for (a) $L/H \gg 1$ and (b) $W/H \gg 1$. }
    \label{fig:6_schem_1dmodel}
\end{figure}
%%%%%

%%%%% Model for large L/H %%%%%
\subsubsection*{Model for large \texorpdfstring{$L/H$}{L/H}}

To discuss the behavior of increasing $L/H$ \tcmagenta{while maintaining $W/H=1$}, we employ a 1D effective model, as shown in Fig.~\ref{fig:6_schem_1dmodel}~(a). 
The slip region propagates along the $y$ direction, as shown in the Supplementary Note and Supplementary Video S1.
Therefore, in this model, the degrees of freedom in the $z$ and $x$ directions are neglected \tcmagenta{by assuming $W/H\ll L/H$}, and the deformation is characterized only by the $y$-dependent displacement in the $y$ direction, $u_y(y, t)$, at the interface $z=0$. 
\tcmagenta{
We also assume a uniform bottom pressure $\Pext$.
}
The equation of motion is given by
\begin{equation}
    \rho \ddot{u}_y(y,t) = \frac{\partial \sigma_{yy}(y,t)}{\partial y} 
    - \frac{\mu( \dot{u}_y(y,t) ) \Pext}{\alphaA H}, 
    \label{eq:eom_1dl}
\end{equation}
where $\ddot{u}_y$ and $\dot{u}_y$ are the second- and first-order time derivatives of the displacement, respectively. 
Here, $\alphaA$ represents the effect of the block thickness and is treated as a fitting parameter. 
The normal stress $\sigma_{yy}$ is given by
\begin{eqnarray}
    \sigma_{yy}(y,t) = E_1 \frac{\partial u_y}{\partial y} + \etat \frac{\partial \dot{u}_y}{\partial y}
    \label{eq:sigma_yy}
\end{eqnarray}
with the elastic constant $E_1 = E/\left\{ (1+\nu)(1-\nu) \right\}$ and viscous constant $\etat = \eta_1 (\eta_1 + 2 \eta_2)/(\eta_1 + \eta_2)$ in the plane stress state \tcmagenta{ by considering the block as a thin plate (see Methods)}.

The quasi-static solution $u^{\mathrm{(a)}}_{y}(y)$ of equation~\eqref{eq:sigma_yy} with $\ddot{u}_y=\dot{u}_y=0$ is obtained analytically, where the precursor slip area $A$ increases with $U$ (see Supplementary Note). 
A linear stability analysis reveals that the quasi-static solution becomes unstable, and bulk sliding occurs when $A$ reaches the critical area $\Ac$ owing to the competition between velocity-weakening friction and viscosity.
The critical area $\Ac$ satisfies
\begin{eqnarray}
    \pi^2 \etat \left( \frac{\Ac}{A_0} \right)^{-2} 
    + 2 \pi L \sqrt{ \rho E_1 } \left( \frac{\Ac}{A_0} \right)^{-1}
    = \frac{\left( \muS - \muK \right) \Pext L^2}{\vc \alphaA H} 
    \label{eq:lc_p_1dl_ver1}
\end{eqnarray}
(see Supplementary Note). 
For $\Ac/A_0 \ll 1$, equation~\eqref{eq:lc_p_1dl_ver1} yields
\begin{equation}
    \frac{\Ac}{A_0} \simeq \pi \left( \frac{\muS - \muK}{\alphaA} \right)^{-\frac{1}{2}}
    \left( \frac{\Pext H}{\etat \vc} \right)^{-\frac{1}{2}}
    \left( \frac{L}{H} \right)^{-1}. 
    \label{eq:lc_p_1dl_ver3}
\end{equation}
This equation indicates that the normalized critical area $\Ac/A_0$ decreases as $L/H$ or $\Pext$ increases, which is consistent with the FEM results shown in Fig.~\ref{fig:3_precursor_slip}~(c).
We plot $\Ac/A_0$ obtained \tcgreen{from} equation~\eqref{eq:lc_p_1dl_ver1} as thin solid lines in Fig.~\ref{fig:3_precursor_slip}~(c) by choosing the fitting parameter $\alphaA = 0.2$ to match the results of the FEM simulations. 
The analytical results semi-quantitatively reproduce the numerical results except for $L/H=1$.

The quasi-static solution $u^{\mathrm{(a)}}_{y}(y)$ yields
\begin{equation}
    \muM = \muK + (\muS - \muK) \frac{\Ac}{A_0}.
    \label{eq:mu_lc_1dl}
\end{equation}
This is consistent with the FEM simulations, as shown by the solid line in Fig.~\ref{fig:4_muM_Ac}.
For $\Ac/A_0 \ll 1$, substituting equation~\eqref{eq:lc_p_1dl_ver3} into this equation, we obtain
\begin{equation}
    \muM - \muK \simeq \pi \left( \muS - \muK \right)^{\frac{1}{2}} \alphaA^{\frac{1}{2}}
    \left( \frac{\Pext H}{\etat \vc} \right)^{-\frac{1}{2}}
    \left( \frac{L}{H} \right)^{-1}. 
    \label{eq:mu_p_1dl}
\end{equation}
This equation indicates that the macroscopic static friction coefficient $\muM$ decreases as $\Pext$ or $L/H$ increases. 
We plot $\muM$ given by equations~\eqref{eq:lc_p_1dl_ver1} and \eqref{eq:mu_lc_1dl} as thin solid lines in Fig.~\ref{fig:2_friction_force}~(b), which semi-quantitatively reproduces the results of the FEM simulations
\tcmagenta{ except for $L/H=1$. }

In a previous study~\cite{Otsuki2013}, $\muM$ is obtained analytically as
\begin{equation}
    \muM - \muK \simeq 
    \frac{\pi^2}{\pi^2 - 4}
    \left( \muS - \muK \right)^{\frac{2}{3}}
    \alpha^{\frac{1}{3}}
    \left( \frac{\Pext H}{\etat \vc} \right)^{-\frac{1}{3}}
    \left( \frac{L}{H} \right)^{-\frac{2}{3}} 
    \label{eq:mu_p_1dl_2013}
\end{equation}
for $\Ac/A_0 \ll 1$ in a system with small $L/H$.
Here, $\alpha$ is the fitting parameter. 
The power-law exponents in equation~\eqref{eq:mu_p_1dl_2013} for the dependence on $\Pext$ and $L/H$ differ from those in equation~\eqref{eq:mu_p_1dl}.
The present model assumes $L/H\gg1$, which results in a uniform bottom pressure, as shown in the Supplementary Note.  
For a small $L/H$, the bottom pressure increases along the driving direction owing to the torque effect~\cite{Otsuki2013, Scheibert2010}, 
\tcmagenta{ 
and the analytical results deviate from those of FEM due to the non-uniform pressure
 as shown in Figs.~\ref{fig:2_friction_force}~(b) and \ref{fig:3_precursor_slip}~(c), 
 } 
 which leads to different exponents from those in the present \tcgreen{study}.

%%%%% Model for large W/H %%%%%
\subsubsection*{Model for large \texorpdfstring{$W/H$}{W/H}}

To discuss the behavior of increasing $W/H$ \tcmagenta{while maintaining $L/H=1$}, we employ a 1D effective model, as shown in Fig.~\ref{fig:6_schem_1dmodel}~(b). 
For $W/H \gg 1$, the slip region propagates along the $x$ direction, as shown in Supplementary Note and Supplementary Video S2.
Hence, in this model, we characterize the deformation only by the $x$-dependent displacement in the $y$ direction, $u_y(x, t)$, at the interface $z=0$ 
\tcmagenta{
by assuming $L/H\ll W/H$. 
We also assume the rod size is sufficiently small and negligible.
}
The equation of motion is given by
\begin{equation}
    \rho \ddot{u}_y(x,t) = 
    \frac{\partial \sigma_{xy}(x,t)}{\partial x}
    - \frac{ \mu(\dot{u}_y(x, t)) \Pext }{ \alphaB H }.
    \label{eq:eom_1dw}
\end{equation}
Here, $\alphaB$ represents the effect of the block thickness and is treated as a fitting parameter.
The shear stress $\sigma_{xy}$ is given by
\begin{eqnarray}
    \sigma_{xy} = E_2 \frac{\partial u_y}{\partial x} 
    + \frac{\eta_1}{2} \frac{\partial \dot{u}_y}{\partial x}
    \label{eq:sigma_xy_1dw}
\end{eqnarray}
with the elastic constant $E_2 = E/\left\{2(1+\nu)\right\}$ and the viscous constant $\eta_1/2$  \tcmagenta{(see Methods)}.

The quasi-static solution $u^{\mathrm{(a)}}_{y}(x)$ is also obtained analytically, where the precursor slip area $A$ increases with the value of $U$ (see Supplementary Note). 
The linear stability analysis reveals that the precursor slip becomes unstable, and bulk sliding occurs when $A$ reaches the critical area $\Ac$ satisfying
\begin{eqnarray}
    2 \pi^2 \eta_1 \left( \frac{\Ac}{A_0} \right)^{-2}
    + 4 \pi W \sqrt{ \rho E_2 } \left( \frac{\Ac}{A_0} \right)^{-1}
    = \frac{\left( \muS - \muK \right) \Pext W^2 }
    { \vc \alphaB H }. 
    \label{eq:lc_p_1dw_ver1}
\end{eqnarray}
For $\Ac/A_0 \ll 1$, this equation yields
\begin{equation}
    \frac{\Ac}{A_0} \simeq \pi \left( \frac{\muS - \muK}{\alphaB} \right)^{-\frac{1}{2}}
    \left( \frac{\Pext H}{2 \eta_1 \vc} \right)^{-\frac{1}{2}}
    \left( \frac{W}{H} \right)^{-1}.
    \label{eq:lc_p_1dw_ver3}
\end{equation}
The power-law exponents for the pressure and aspect ratio are the same as those in equation~\eqref{eq:lc_p_1dl_ver3}.
This equation indicates that $\Ac/A_0$ decreases as $\Pext$ or $W/H$ increases. 
We plot $\Ac/A_0$ given by equation~\eqref{eq:lc_p_1dw_ver1} as thin solid lines in Fig.~\ref{fig:3_precursor_slip}~(d), which semi-quantitatively reproduces the results of the FEM analysis by choosing $\alphaB=0.2$ 
\tcmagenta{
except for $W/H=1$.
For small $W/H$, the size of the rod and the $y$-dependence of the displacement become relevant, which leads to the deviation between the numerical and theoretical results.
}

The macroscopic static friction coefficient $\muM$ is given by equation~\eqref{eq:mu_lc_1dl}. 
For $\Ac/A_0 \ll 1$, substituting equation~\eqref{eq:lc_p_1dw_ver3} into equation~\eqref{eq:mu_lc_1dl}, we obtain
\begin{equation}
    \muM - \muK \simeq \pi \left( \muS - \muK \right)^{\frac{1}{2}} \alphaB^{\frac{1}{2}}
    \left( \frac{\Pext H}{2 \eta_1 \vc} \right)^{-\frac{1}{2}}
    \left( \frac{W}{H} \right)^{-1}.
    \label{eq:mu_p_1dw}
\end{equation}
The macroscopic static friction coefficient $\muM$ decreases as $\Pext$ or $W/H$ increases.
The thin solid lines shown in Fig.~\ref{fig:2_friction_force}~(c) are given by equations~\eqref{eq:mu_lc_1dl} and \eqref{eq:lc_p_1dw_ver1}, and \tcgreen{they} semi-quantitatively reproduce the results of the FEM simulations
\tcmagenta{
except for $W/H=1$. 
}

%%%%%%%%%%%%%%%%%%%%%%
%%%%% Discussion %%%%%
%%%%%%%%%%%%%%%%%%%%%%
\section*{Discussion}
%%%%%%%%%%%%%%%%%%%%%%

In this study, we numerically investigate the sliding motion of a 3D \tcgreen{viscoelastic} object using the FEM. 
The critical area of the precursor slip and macroscopic static friction coefficient \tcgreen{decrease} with an increase in the external pressure, length, or width of the object. 
The analysis \tcgreen{based on} the simplified models \tcgreen{reveals} that the stability condition determines the critical area of the precursor slip owing to the competition between the velocity-weakening friction and viscosity. 
The analysis explains the dependence of macroscopic static friction in the FEM simulations.

In a previous study~\cite{Otsuki2013}, the aspect ratio of the system is fixed at $L/H=2$ to investigate the size and load dependences of the precursor slip and the breakdown of Amontons' law. 
For $L/H=2$, the nonuniformity of the bottom pressure is remarkable, which \tcgreen{is} considered to be the origin of the precursor slip and the breakdown of Amontons' law. 
However, the present results with various aspect ratios show that the nonuniformity of shear stress also causes these behaviors without non-uniform pressure. 
Although the model \tcgreen{considered in} the previous study reproduces the results of systems with a smaller $L/H$ better, the simplified model in this study is more appropriate for systems with a large $L/H$ (see Supplementary Note).

The parameters for the FEM simulations employed here are those of a virtual material, and different from those \tcgreen{of poly methyl methacrylate (PMMA) employed in experiments}~\cite{Rubinstein2007, Katano2014}.
We choose them to compare our results with the 2D simulations of previous studies~\cite{Otsuki2013} and to reduce the computational load (see Methods). 
\tcmagenta{
It also should be noted that the driving rod employed in experiments is hard but has finite stiffness, which is different from the rigid rod used in this study. 
The effect of the finite stiffness of the driving rod is considered to be small because it is taken into account as a deformation of the viscoelastic block around the driving point. 
In addition, we have ignored the aging effect~\cite{Baumberger2006} in the local friction model
because a previous experiment using PMMA~\cite{Rubinstein2007} indicates that the time scale of the aging is larger than that of the stick of the macroscopic stick-slip motion. 
The difference in the parameters, the driving method, and the local friction model may affect our results.
However, FEM simulations employing similar parameters semi-quantitatively reproduce the external pressure dependence of the macroscopic static friction coefficient obtained in the \tcgreen{experiment using PMMA}~\cite{Katano2014}. 
}
The dependence of the macroscopic static friction coefficient on the aspect ratio for PMMA is also considered to be consistent with our present results. 
\tcmagenta{
The dependence on material parameters, the driving methods, and the local friction model will be investigated in future work. 
}

\tcmagenta{
The dependence of the static and kinetic friction coefficients on the pressure or block shape has been studied in experiments using rubber blocks~\cite{Maegawa2016, Moriyasu2019, Hale2020}. 
The results of these experiments are partially consistent with ours, but there is a difference in the dependence on the aspect ratio. 
In these experiments, the methods to change the aspect ratio and drive the block differ from those used in this study.
For the rubber block, the local Amontons' law used in this study may not be applicable because the real contact area can become comparable to the apparent contact area, which contradicts the assumption of the Amontons' law. 
We need further investigations to determine the origin of the difference.
}

Recent numerical simulations of spring-block models have shown that the friction coefficient changes with the geometric pattern of the frictional interface~\cite{Costagliola2016, Maegawa2017, Costagliola2018, Costagliola2022}. 
However, our results indicate that an object shape can also control the macroscopic static friction coefficient. 
This might lead to new insights into methods for controlling friction in various \tcgreen{objects,} including shoe soles and tires.

Precursor slip has been investigated experimentally for the sliding motion of PMMA blocks based on fracture mechanics~\cite{Kammer2015, Svetlizky2014, Bayart2016, Svetlizky2017, Berman2020, Gvirtzman2021}. 
Such a precursor slip is related to pre-earthquakes that occur a few days or months before a major earthquake~\cite{Kato2012, Obara2016, Kato2021}, which are studied using frictional spring-block models~\cite{Petrillo2020}. 
However, these studies have focused on 1D frictional interfaces or discrete models, which differ from 2D friction interfaces in more realistic systems.
Our results for a 3D system with a 2D interface will provide new insights into \tcgreen{the} precursor slip \tcgreen{observed} in realistic situations.

%%%%%%%%%%%%%%%%%%%
%%%%% Methods %%%%%
%%%%%%%%%%%%%%%%%%%
\section*{Methods}
%%%%%%%%%%%%%%%%%%%

%%%%% Setting of system %%%%%
\subsection*{Setting of system}

The equation of motion for a viscoelastic body is given by 
\begin{equation}
    \rho \ddot{\bm{u}} = \bm{\nabla} \cdot \bm{\sigma}
    \label{eq:eom_block}
\end{equation}
with displacement $\bm{u}$, stress $\bm{\sigma}$, and second-order time derivative $\ddot{\bm{u}}$ of displacement. 
The stress $\bm{\sigma}$ is given by the sum of the elastic stress $\bm{\sigma}^\mathrm{(E)}$ obeying Hooke's law and the viscous stress $\bm{\sigma}^\mathrm{(V)}$, which is proportional to the strain rate. 
We assume that the viscoelastic body is isotropic. 
The elastic stress tensor $\sigma^\mathrm{(E)}_{ij}$ is given by
\begin{equation}
    \sigma^\mathrm{(E)}_{ij} = \frac{E}{1+\nu} \,\epsilon_{ij} + \frac{\nu E}{(1+\nu)(1-2\nu)} \,\epsilon_{kk} \delta_{ij}
    \label{eq:sigma_E}
\end{equation}
with the Kronecker delta $\delta_{ij}$ and the strain tensor $\epsilon_{ij}$. 
The viscous stress tensor $\sigma^{ ( \mathrm{V} ) }_{ij}$ is given by
\begin{equation}
    \sigma^\mathrm{(V)}_{ij} = \eta_1\,\Dot{\epsilon}_{ij} + \eta_2\,\Dot{\epsilon}_{kk} \delta_{ij}
    \label{eq:sigma_V}
\end{equation}
with the strain rate tensor $\Dot{\epsilon}_{ij}$~\cite{Landau1986}. 
%%%%%
The boundary conditions for the top surface at $z=H$ are $\sigma_{zz} = -P_{\mathrm{ext}}$ and $\sigma_{zx} = \sigma_{zy} = 0$.
At the free surface for $x=0,W$ or $y=0,L$, we assume $\bm{\sigma} \cdot \bm{n} = \bm{0}$ with the normal vector $\bm{n}$ of the surface. 
The boundary \tcgreen{conditions} at the contact surface with a rigid rod ($y=0$) \tcgreen{are} given by $\sigma_{yx} = \sigma_{yz} = 0$ and $\dot{u}_y = \Vrod$, where $\dot{u}_y$ is the velocity in the $y$ direction \tcmagenta{and $\Vrod$ is the velocity of rigid rod}. 
%%%%%
At the bottom of the block ($z=0$) in contact with a rigid substrate, the bottom pressure $p = -\sigma_{zz}$ is determined such that the displacement $u_z$ in the $z$ direction is $0$. 
However, the bottom pressure is limited to $p \geq 0$. 
The region of the bottom surface with $u_z > 0$ and $p=0$ becomes a free surface with $\bm{\sigma} \cdot \bm{n} = \bm{0}$. 
The boundary condition in the tangential direction at the bottom with $p>0$ is given by
\begin{equation}
    \bm{t} = -\bm{v}/v ~ \mu(v)p
    \label{eq:vector_t}
\end{equation}
with the tangential stress vector $\bm{t}(x,y) = (\sigma_{zx},\sigma_{zy})$, local slip velocity vector $\bm{v}(x,y) = (\dot{u}_x, \dot{u}_y)$, velocity $\dot{u}_x$ in the $x$ direction, and velocity $\dot{u}_y$ in the $y$ direction. 
\tcmagenta{The direction of the frictional stress is opposite to that of the local slip velocity. }
Frictional stress is defined as $\sigmaf(x,y) = |\bm{t}|$. 
The slip velocity is defined as $v(x,y) = |\bm{v}(x,y) |$. 

\tcmagenta{
The frictional stress $\sigmaf$ is given by equation~\eqref{eq:sigma_fric}. In the case $v(x,y) = 0$, the frictional stress is balanced with the local shear stress, where the maximum magnitude of the former is given by $\muS p(x, y)$. 
The local friction coefficient $\mu(v)$ linearly decreases from $\muS$ to $\muK$ for $0 < v \leq \vc$ and $\muK$ for $v > \vc$. 
Amontons' law is expected to hold locally if the local region considered in the frictional interface contains a sufficiently large number of real contact points and has negligibly small spatial variations in internal stress~\cite{Archard1957, Dieterich1996, Dieterich1994}. 
}

To treat static friction in the numerical simulation, we introduce a small velocity scale $v_\mathrm{e}$. 
The local friction coefficient $\mu(v)$ is given by
\begin{equation}
    \mu (v) = \left\{
    \begin{array}{lll}
        \muS\,v/v_\mathrm{e}, ~& 0 \leq v \leq v_\mathrm{e} \\
        \muS - \left( \muS -\muK \right)v/\vc, ~& 
        v_\mathrm{e} < v < \vc \\
        \muK, ~& v \geq \vc
    \end{array}
    \right. ~.
    \label{eq:mu_v}
\end{equation}
We consider the limit $v_\mathrm{e} \rightarrow +0$.
The region with $0 \leq v \leq v_\mathrm{e}$ corresponds to static friction.
The slip area $A$ is defined as the region with $v > v_\mathrm{e}$.

%%%%% 3D FEM simulation %%%%%
\subsection*{\tcmagenta{Details of 3D FEM simulation}}

The viscoelastic block is divided into cubes \tcgreen{with} length $\Delta x$ consisting of six tetrahedra. 
The displacements and velocities within each element are approximated using a linear interpolation. 
We choose the characteristic velocity $v_\mathrm{e}/\Vrod = 2.5\times10 ^{-2}$ such that $v_\mathrm{e}/\Vrod \ll 1$ is satisfied. 
\tcmagenta{
In the FEM simulations, we select $\Delta x/H=1/40$, $\Delta t/( H \sqrt{\rho/E} )\thickapprox 10^{-6}$, and $\Vrod \sqrt{\rho/E}=2\times 10^{-5}$.
We have confirmed that the numerical results do not change, even if we use smaller values. 
}

\tcmagenta{
First, we apply an external uniform pressure $\Pext$ to the top surface and relax the system to an equilibrium state.
After relaxation, the center of the side surface $(x,y,z)=(W/2, 0, H/2)$ is pushed along the $y$ direction by a rigid rod from time $t=0$ with a sufficiently slow speed $\Vrod$. 
The displacement of the rigid rod is denoted by $U(t)= \Vrod t$. 
The length of one side of a rigid square rod is $0.1H$, and the height of its center from the bottom is $0.5H$. 
}

%%%%% Model analysis %%%%%
\subsection*{\tcmagenta{Details of analysis based on simplified models}}

\tcmagenta{
\noindent\textbf{Model for large $L/H$}: 
The second term on the right-hand side of equation~\eqref{eq:eom_1dl} represents local friction. 
Here, we assume a constant bottom pressure given by $\Pext$, which is verified in the FEM simulations for $L/H \gg 1$ as shown in the Supplementary Note and Supplementary Video S1.
The local friction coefficient $\mu$ is expressed as a function of $v=|\dot{u}_y|$. 
Note that $0 \le \mu \le \muS$ when $v=0$. 
The boundary conditions are $\partial u_y(y=L, t)/\partial y = 0$ and $u_y(y=0, t) = U(t)$. 
In our analysis, we set the origin of $U$ immediately after the bulk sliding and \tcgreen{assume} that the ratio of the frictional stress to $\Pext$ is equal to $\muK$ at $U=0$. 
}

\noindent
\tcmagenta{
\textbf{Model for large $W/H$}: 
The second term on the right-hand side of equation~\eqref{eq:eom_1dw} represents the friction. 
The bottom pressure is almost independent of $x$ in the FEM simulations, as shown in the Supplementary Note and Supplementary Video S2. Therefore, we assume a constant bottom pressure given by $\Pext$. 
The boundary conditions are $\partial u_y (|x|=W/2, t)/\partial x = 0$ and $u_y(x=0, t) = U(t)$. 
}

%%%%% Parameters %%%%%
\subsection*{Parameters}

The parameters for the viscoelastic object are chosen as $\nu=0.34$, $\eta_1/( H \sqrt{\rho E} ) = 2$, and $\eta_2/\eta_1=1$, whereas we set the parameters for the friction as $\muS=0.38$, $\muK=0.1$, and $\vc \sqrt{\rho / E} = 3.4\times10^{-4}$\tcgreen{,} following previous FEM simulations~\cite{Otsuki2013}. 
\tcmagenta{
These values are different from those adopted for the experiment using PMMA~\cite{Otsuki2013, Katano2014}. 
}
The parameters for the PMMA blocks~\cite{Katano2014} are estimated as $L/H=5$, $W/H = 0.25$, $\Pext/E \thickapprox 3 \times 10^{-4}$, $\nu = 0.4$, $\muS = 1.2$, and $\muK = 0.2$, and \tcgreen{much} smaller $\vc \sqrt{\rho/E }$ and $\eta_1 / (H \sqrt{\rho E})$ are used in \tcgreen{the} previous study~\cite{Otsuki2013}.

%%%%% Data availability %%%%%
\section*{Data availability}
The datasets used and/or analyzed during the current study available from the corresponding author on reasonable request. 

%%%%%%%%%%%%%%%
%%%%% bib %%%%%
%%%%%%%%%%%%%%%
\bibliography{bib_imo2022}

\begin{thebibliography}{10}
\urlstyle{rm}
\expandafter\ifx\csname url\endcsname\relax
  \def\url#1{\texttt{#1}}\fi
\expandafter\ifx\csname urlprefix\endcsname\relax\def\urlprefix{URL }\fi
\expandafter\ifx\csname doiprefix\endcsname\relax\def\doiprefix{DOI: }\fi
\providecommand{\bibinfo}[2]{#2}
\providecommand{\eprint}[2][]{\url{#2}}

\bibitem{Bowden1950}
\bibinfo{author}{Bowden, F.~P.} \& \bibinfo{author}{Tabor, D.}
\newblock \emph{\bibinfo{title}{{The Friction and Lubrication of Solids}}}
  (\bibinfo{publisher}{Oxford University Press}, \bibinfo{address}{New York},
  \bibinfo{year}{1950}).

\bibitem{Persson2000}
\bibinfo{author}{Persson, B. N.~J.}
\newblock \emph{\bibinfo{title}{{Sliding Friction: Physical Principles and
  Applications}}} (\bibinfo{publisher}{Springer}, \bibinfo{address}{Berlin},
  \bibinfo{year}{2000}), \bibinfo{edition}{2} edn.

\bibitem{Popov2017_text}
\bibinfo{author}{Popov, V.~L.}
\newblock \emph{\bibinfo{title}{{Contact Mechanics and Friction: Physical
  Principles and Applications}}} (\bibinfo{publisher}{Springer},
  \bibinfo{address}{Berlin}, \bibinfo{year}{2017}), \bibinfo{edition}{2} edn.

\bibitem{Rabinowicz1995}
\bibinfo{author}{Rabinowicz, E.}
\newblock \emph{\bibinfo{title}{{Friction and Wear of Materials}}}
  (\bibinfo{publisher}{John Wiley \& Sons}, \bibinfo{address}{New York},
  \bibinfo{year}{1995}), \bibinfo{edition}{2} edn.

\bibitem{Dowson1998}
\bibinfo{author}{Dowson, D.}
\newblock \emph{\bibinfo{title}{{History of Tribology}}}
  (\bibinfo{publisher}{John Wiley \& Sons}, \bibinfo{address}{New York},
  \bibinfo{year}{1998}), \bibinfo{edition}{2} edn.

\bibitem{Bhushan2013}
\bibinfo{author}{Bhushan, B.}
\newblock \emph{\bibinfo{title}{{Principles and Applications of Tribology}}}
  (\bibinfo{publisher}{John Wiley \& Sons}, \bibinfo{address}{New York},
  \bibinfo{year}{2013}), \bibinfo{edition}{2} edn.

\bibitem{Baumberger2006}
\bibinfo{author}{Baumberger, T.} \& \bibinfo{author}{Caroli, C.}
\newblock \bibinfo{journal}{\bibinfo{title}{{Solid friction from stick-slip
  down to pinning and aging}}}.
\newblock {\emph{\JournalTitle{Advances in Physics}}}
  \textbf{\bibinfo{volume}{55}}, \bibinfo{pages}{279--348},
  \doiprefix\url{https://doi.org/10.1080/00018730600732186}
  (\bibinfo{year}{2006}).

\bibitem{Archard1957}
\bibinfo{author}{Archard, J.~F.}
\newblock \bibinfo{journal}{\bibinfo{title}{{Elastic deformation and the laws
  of friction}}}.
\newblock {\emph{\JournalTitle{Proceedings of the Royal Society of London A}}}
  \textbf{\bibinfo{volume}{243}}, \bibinfo{pages}{190--205},
  \doiprefix\url{https://doi.org/10.1098/rspa.1957.0214}
  (\bibinfo{year}{1957}).

\bibitem{Dieterich1996}
\bibinfo{author}{Dieterich, J.~H.} \& \bibinfo{author}{Kilgore, B.~D.}
\newblock \bibinfo{journal}{\bibinfo{title}{{Imaging surface contacts: power
  law contact distributions and contact stresses in quartz, calcite, glass and
  acrylic plastic}}}.
\newblock {\emph{\JournalTitle{Tectonophysics}}}
  \textbf{\bibinfo{volume}{256}}, \bibinfo{pages}{219--239},
  \doiprefix\url{https://doi.org/10.1016/0040-1951(95)00165-4}
  (\bibinfo{year}{1996}).

\bibitem{Otsuki2013}
\bibinfo{author}{Otsuki, M.} \& \bibinfo{author}{Matsukawa, H.}
\newblock \bibinfo{journal}{\bibinfo{title}{{Systematic breakdown of Amontons'
  law of friction for an elastic object locally obeying Amontons' law}}}.
\newblock {\emph{\JournalTitle{Scientific Reports}}}
  \textbf{\bibinfo{volume}{3}}, \bibinfo{pages}{1586},
  \doiprefix\url{https://doi.org/10.1038/srep01586} (\bibinfo{year}{2013}).

\bibitem{Ozaki2014}
\bibinfo{author}{Ozaki, S.}, \bibinfo{author}{Inanobe, C.} \&
  \bibinfo{author}{Nakano, K.}
\newblock \bibinfo{journal}{\bibinfo{title}{{Finite element analysis of
  precursors to macroscopic stick-slip motion in elastic materials: analysis of
  friction test as a boundary value problem}}}.
\newblock {\emph{\JournalTitle{Tribology Letters}}}
  \textbf{\bibinfo{volume}{55}}, \bibinfo{pages}{151--163},
  \doiprefix\url{https://doi.org/10.1007/s11249-014-0343-y}
  (\bibinfo{year}{2014}).

\bibitem{Katano2014}
\bibinfo{author}{Katano, Y.}, \bibinfo{author}{Nakano, K.},
  \bibinfo{author}{Otsuki, M.} \& \bibinfo{author}{Matsukawa, H.}
\newblock \bibinfo{journal}{\bibinfo{title}{{Novel friction law for the static
  friction force based on local precursor slipping}}}.
\newblock {\emph{\JournalTitle{Scientific Reports}}}
  \textbf{\bibinfo{volume}{4}}, \bibinfo{pages}{6324},
  \doiprefix\url{https://doi.org/10.1038/srep06324} (\bibinfo{year}{2014}).

\bibitem{Bouissou1998}
\bibinfo{author}{Bouissou, S.}, \bibinfo{author}{Petit, J.~P.} \&
  \bibinfo{author}{Barquins, M.}
\newblock \bibinfo{journal}{\bibinfo{title}{{Normal load, slip rate and
  roughness influence on the polymethylmethacrylate dynamics of sliding 1.
  Stable sliding to stick-slip transition}}}.
\newblock {\emph{\JournalTitle{Wear}}} \textbf{\bibinfo{volume}{214}},
  \bibinfo{pages}{156--164},
  \doiprefix\url{https://doi.org/10.1016/S0043-1648(97)00242-1}
  (\bibinfo{year}{1998}).

\bibitem{Rubinstein2004}
\bibinfo{author}{Rubinstein, S.~M.}, \bibinfo{author}{Cohen, G.} \&
  \bibinfo{author}{Fineberg, J.}
\newblock \bibinfo{journal}{\bibinfo{title}{{Detachment fronts and the onset of
  dynamic friction}}}.
\newblock {\emph{\JournalTitle{Nature}}} \textbf{\bibinfo{volume}{430}},
  \bibinfo{pages}{1005--1009},
  \doiprefix\url{https://doi.org/10.1038/nature02830} (\bibinfo{year}{2004}).

\bibitem{Rubinstein2007}
\bibinfo{author}{Rubinstein, S.~M.}, \bibinfo{author}{Cohen, G.} \&
  \bibinfo{author}{Fineberg, J.}
\newblock \bibinfo{journal}{\bibinfo{title}{{Dynamics of precursors to
  frictional sliding}}}.
\newblock {\emph{\JournalTitle{Physical Review Letters}}}
  \textbf{\bibinfo{volume}{98}}, \bibinfo{pages}{226103},
  \doiprefix\url{https://doi.org/10.1103/PhysRevLett.98.226103}
  (\bibinfo{year}{2007}).

\bibitem{Ben-David2010}
\bibinfo{author}{Ben-David, O.}, \bibinfo{author}{Cohen, G.} \&
  \bibinfo{author}{Fineberg, J.}
\newblock \bibinfo{journal}{\bibinfo{title}{{The dynamics of the onset of
  frictional slip}}}.
\newblock {\emph{\JournalTitle{Science}}} \textbf{\bibinfo{volume}{330}},
  \bibinfo{pages}{211--214},
  \doiprefix\url{https://www.science.org/doi/abs/10.1126/science.1194777}
  (\bibinfo{year}{2010}).

\bibitem{Ben-David2011}
\bibinfo{author}{Ben-David, O.} \& \bibinfo{author}{Fineberg, J.}
\newblock \bibinfo{journal}{\bibinfo{title}{{Static friction coefficient is not
  a material constant}}}.
\newblock {\emph{\JournalTitle{Physical Review Letters}}}
  \textbf{\bibinfo{volume}{106}}, \bibinfo{pages}{254301},
  \doiprefix\url{https://doi.org/10.1103/PhysRevLett.106.254301}
  (\bibinfo{year}{2011}).

\bibitem{Malthe-Sorenssen2021}
\bibinfo{author}{Malthe-S{\o}renssen, A.}
\newblock \bibinfo{journal}{\bibinfo{title}{{The onset of a slip}}}.
\newblock {\emph{\JournalTitle{Nature Physics}}} \textbf{\bibinfo{volume}{17}},
  \bibinfo{pages}{983--985},
  \doiprefix\url{https://doi.org/10.1038/s41567-021-01312-1}
  (\bibinfo{year}{2021}).

\bibitem{Braun2009}
\bibinfo{author}{Braun, O.~M.}, \bibinfo{author}{Barel, I.} \&
  \bibinfo{author}{Urbakh, M.}
\newblock \bibinfo{journal}{\bibinfo{title}{{Dynamics of transition from static
  to kinetic friction}}}.
\newblock {\emph{\JournalTitle{Physical Review Letters}}}
  \textbf{\bibinfo{volume}{103}}, \bibinfo{pages}{194301},
  \doiprefix\url{https://doi.org/10.1103/PhysRevLett.103.194301}
  (\bibinfo{year}{2009}).

\bibitem{Maegawa2010}
\bibinfo{author}{Maegawa, S.}, \bibinfo{author}{Suzuki, A.} \&
  \bibinfo{author}{Nakano, K.}
\newblock \bibinfo{journal}{\bibinfo{title}{{Precursors of global slip in a
  longitudinal line contact under non-uniform normal loading}}}.
\newblock {\emph{\JournalTitle{Tribology Letters}}}
  \textbf{\bibinfo{volume}{38}}, \bibinfo{pages}{313--323},
  \doiprefix\url{https://doi.org/10.1007/s11249-010-9611-7}
  (\bibinfo{year}{2010}).

\bibitem{Scheibert2010}
\bibinfo{author}{Scheibert, J.} \& \bibinfo{author}{Dysthe, D.~K.}
\newblock \bibinfo{journal}{\bibinfo{title}{{Role of friction-induced torque in
  stick-slip motion}}}.
\newblock {\emph{\JournalTitle{Europhysics Letters}}}
  \textbf{\bibinfo{volume}{92}}, \bibinfo{pages}{54001},
  \doiprefix\url{https://doi.org/10.1209/0295-5075/92/54001}
  (\bibinfo{year}{2010}).

\bibitem{Amundsen2012}
\bibinfo{author}{Amundsen, D.~S.}, \bibinfo{author}{Scheibert, J.},
  \bibinfo{author}{Th{\o}gersen, K.}, \bibinfo{author}{Tr{\o}mborg, J.} \&
  \bibinfo{author}{Malthe-S{\o}renssen, A.}
\newblock \bibinfo{journal}{\bibinfo{title}{{1D model of precursors to
  frictional stick-slip motion allowing for robust comparison with
  experiments}}}.
\newblock {\emph{\JournalTitle{Tribology Letters}}}
  \textbf{\bibinfo{volume}{45}}, \bibinfo{pages}{357--369},
  \doiprefix\url{https://doi.org/10.1007/s11249-011-9894-3}
  (\bibinfo{year}{2012}).

\bibitem{Tromborg2011}
\bibinfo{author}{Tr{\o}mborg, J.}, \bibinfo{author}{Scheibert, J.},
  \bibinfo{author}{Amundsen, D.~S.}, \bibinfo{author}{Th{\o}gersen, K.} \&
  \bibinfo{author}{Malthe-S{\o}renssen, A.}
\newblock \bibinfo{journal}{\bibinfo{title}{{Transition from static to kinetic
  friction: insights from a 2D model}}}.
\newblock {\emph{\JournalTitle{Physical Review Letters}}}
  \textbf{\bibinfo{volume}{107}}, \bibinfo{pages}{074301},
  \doiprefix\url{https://doi.org/10.1103/PhysRevLett.107.074301}
  (\bibinfo{year}{2011}).

\bibitem{Tromborg2014}
\bibinfo{author}{Tr{\o}mborg, J.~K.} \emph{et~al.}
\newblock \bibinfo{journal}{\bibinfo{title}{{Slow slip and the transition from
  fast to slow fronts in the rupture of frictional interfaces}}}.
\newblock {\emph{\JournalTitle{Proceedings of the National Academy of Sciences
  of the U.S.A.}}} \textbf{\bibinfo{volume}{111}}, \bibinfo{pages}{8764--8769},
  \doiprefix\url{https://doi.org/10.1073/pnas.1321752111}
  (\bibinfo{year}{2014}).

\bibitem{Radiguet2013}
\bibinfo{author}{Radiguet, M.}, \bibinfo{author}{Kammer, D.~S.},
  \bibinfo{author}{Gillet, P.} \& \bibinfo{author}{Molinari, J.-F.}
\newblock \bibinfo{journal}{\bibinfo{title}{{Survival of heterogeneous stress
  distributions created by precursory slip at frictional interfaces}}}.
\newblock {\emph{\JournalTitle{Physical Review Letters}}}
  \textbf{\bibinfo{volume}{111}}, \bibinfo{pages}{164302},
  \doiprefix\url{https://doi.org/10.1103/PhysRevLett.111.164302}
  (\bibinfo{year}{2013}).

\bibitem{Kammer2015}
\bibinfo{author}{Kammer, D.~S.}, \bibinfo{author}{Radiguet, M.},
  \bibinfo{author}{Ampuero, J.-P.} \& \bibinfo{author}{Molinari, J.-F.}
\newblock \bibinfo{journal}{\bibinfo{title}{{Linear elastic fracture mechanics
  predicts the propagation distance of frictional slip}}}.
\newblock {\emph{\JournalTitle{Tribology Letters}}}
  \textbf{\bibinfo{volume}{57}}, \bibinfo{pages}{23},
  \doiprefix\url{https://doi.org/10.1007/s11249-014-0451-8}
  (\bibinfo{year}{2015}).

\bibitem{Taloni2015}
\bibinfo{author}{Taloni, A.}, \bibinfo{author}{Benassi, A.},
  \bibinfo{author}{Sandfeld, S.} \& \bibinfo{author}{Zapperi, S.}
\newblock \bibinfo{journal}{\bibinfo{title}{{Scalar model for frictional
  precursors dynamics}}}.
\newblock {\emph{\JournalTitle{Scientific Reports}}}
  \textbf{\bibinfo{volume}{5}}, \bibinfo{pages}{8086},
  \doiprefix\url{https://doi.org/10.1038/srep08086} (\bibinfo{year}{2015}).

\bibitem{deGeus2019}
\bibinfo{author}{de~Geus, T. W.~J.}, \bibinfo{author}{Popović, M.},
  \bibinfo{author}{Ji, W.}, \bibinfo{author}{Rosso, A.} \&
  \bibinfo{author}{Wyart, M.}
\newblock \bibinfo{journal}{\bibinfo{title}{How collective asperity detachments
  nucleate slip at frictional interfaces}}.
\newblock {\emph{\JournalTitle{Proceedings of the National Academy of Sciences
  of the U.S.A.}}} \textbf{\bibinfo{volume}{116}},
  \bibinfo{pages}{23977--23983},
  \doiprefix\url{https://doi.org/10.1073/pnas.1906551116}
  (\bibinfo{year}{2019}).

\bibitem{ccm2006}
\bibinfo{author}{Wriggers, P.}
\newblock \emph{\bibinfo{title}{{Computational Contact Mechanics}}}
  (\bibinfo{publisher}{Springer}, \bibinfo{address}{Berlin},
  \bibinfo{year}{2006}), \bibinfo{edition}{2} edn.

\bibitem{Maegawa2016}
\bibinfo{author}{Maegawa, S.}, \bibinfo{author}{Itoigawa, F.} \&
  \bibinfo{author}{Nakamura, T.}
\newblock \bibinfo{journal}{\bibinfo{title}{A role of friction-induced torque
  in sliding friction of rubber materials}}.
\newblock {\emph{\JournalTitle{Tribology International}}}
  \textbf{\bibinfo{volume}{93}}, \bibinfo{pages}{182--189},
  \doiprefix\url{https://doi.org/10.1016/j.triboint.2015.08.030}
  (\bibinfo{year}{2016}).

\bibitem{Moriyasu2019}
\bibinfo{author}{Moriyasu, K.}, \bibinfo{author}{Nishiwaki, T.},
  \bibinfo{author}{Shibata, K.}, \bibinfo{author}{Yamaguchi, T.} \&
  \bibinfo{author}{Hokkirigawa, K.}
\newblock \bibinfo{journal}{\bibinfo{title}{Friction control of a resin
  foam/rubber laminated block material}}.
\newblock {\emph{\JournalTitle{Tribology International}}}
  \textbf{\bibinfo{volume}{136}}, \bibinfo{pages}{548--555},
  \doiprefix\url{https://doi.org/10.1016/j.triboint.2019.04.024}
  (\bibinfo{year}{2019}).

\bibitem{Hale2020}
\bibinfo{author}{Hale, J.}, \bibinfo{author}{Lewis, R.} \&
  \bibinfo{author}{Carré, M.~J.}
\newblock \bibinfo{journal}{\bibinfo{title}{Rubber friction and the effect of
  shape}}.
\newblock {\emph{\JournalTitle{Tribology International}}}
  \textbf{\bibinfo{volume}{141}}, \bibinfo{pages}{105911},
  \doiprefix\url{https://doi.org/10.1016/j.triboint.2019.105911}
  (\bibinfo{year}{2020}).

\bibitem{Costagliola2016}
\bibinfo{author}{Costagliola, G.}, \bibinfo{author}{Bosia, F.} \&
  \bibinfo{author}{Pugno, N.~M.}
\newblock \bibinfo{journal}{\bibinfo{title}{{Static and dynamic friction of
  hierarchical surfaces}}}.
\newblock {\emph{\JournalTitle{Physical Review E}}}
  \textbf{\bibinfo{volume}{94}}, \bibinfo{pages}{063003},
  \doiprefix\url{https://doi.org/10.1103/PhysRevE.94.063003}
  (\bibinfo{year}{2016}).

\bibitem{Maegawa2017}
\bibinfo{author}{Maegawa, S.}, \bibinfo{author}{Itoigawa, F.},
  \bibinfo{author}{Nakamura, T.}, \bibinfo{author}{Matsuoka, H.} \&
  \bibinfo{author}{Fukui, S.}
\newblock \bibinfo{journal}{\bibinfo{title}{{Effect of tangential loading
  history on static friction force of elastic slider with split contact
  surface: model calculation}}}.
\newblock {\emph{\JournalTitle{Tribology Letters}}}
  \textbf{\bibinfo{volume}{65}}, \bibinfo{pages}{37},
  \doiprefix\url{https://doi.org/10.1007/s11249-017-0811-2}
  (\bibinfo{year}{2017}).

\bibitem{Costagliola2018}
\bibinfo{author}{Costagliola, G.}, \bibinfo{author}{Bosia, F.} \&
  \bibinfo{author}{Pugno, N.~M.}
\newblock \bibinfo{journal}{\bibinfo{title}{{A 2-D model for friction of
  complex anisotropic surfaces}}}.
\newblock {\emph{\JournalTitle{Journal of the Mechanics and Physics of
  Solids}}} \textbf{\bibinfo{volume}{112}}, \bibinfo{pages}{50--65},
  \doiprefix\url{https://doi.org/10.1016/j.jmps.2017.11.015}
  (\bibinfo{year}{2018}).

\bibitem{Costagliola2022}
\bibinfo{author}{Costagliola, G.}, \bibinfo{author}{Bosia, F.} \&
  \bibinfo{author}{Pugno, N.~M.}
\newblock \bibinfo{journal}{\bibinfo{title}{{Correlation between slip
  precursors and topological length scales at the onset of frictional
  sliding}}}.
\newblock {\emph{\JournalTitle{International Journal of Solids and
  Structures}}} \textbf{\bibinfo{volume}{243}}, \bibinfo{pages}{111525},
  \doiprefix\url{https://doi.org/10.1016/j.ijsolstr.2022.111525}
  (\bibinfo{year}{2022}).

\bibitem{Svetlizky2014}
\bibinfo{author}{Svetlizky, I.} \& \bibinfo{author}{Fineberg, J.}
\newblock \bibinfo{journal}{\bibinfo{title}{{Classical shear cracks drive the
  onset of dry frictional motion}}}.
\newblock {\emph{\JournalTitle{Nature}}} \textbf{\bibinfo{volume}{509}},
  \bibinfo{pages}{205--208},
  \doiprefix\url{https://doi.org/10.1038/nature13202} (\bibinfo{year}{2014}).

\bibitem{Bayart2016}
\bibinfo{author}{Bayart, E.}, \bibinfo{author}{Svetlizky, I.} \&
  \bibinfo{author}{Fineberg, J.}
\newblock \bibinfo{journal}{\bibinfo{title}{Fracture mechanics determine the
  lengths of interface ruptures that mediate frictional motion}}.
\newblock {\emph{\JournalTitle{Nature Physics}}} \textbf{\bibinfo{volume}{12}},
  \bibinfo{pages}{166--170}, \doiprefix\url{https://doi.org/10.1038/nphys3539}
  (\bibinfo{year}{2016}).

\bibitem{Svetlizky2017}
\bibinfo{author}{Svetlizky, I.}, \bibinfo{author}{Kammer, D.~S.},
  \bibinfo{author}{Bayart, E.}, \bibinfo{author}{Cohen, G.} \&
  \bibinfo{author}{Fineberg, J.}
\newblock \bibinfo{journal}{\bibinfo{title}{{Brittle fracture theory predicts
  the equation of motion of frictional rupture fronts}}}.
\newblock {\emph{\JournalTitle{Physical Review Letters}}}
  \textbf{\bibinfo{volume}{118}}, \bibinfo{pages}{125501},
  \doiprefix\url{https://doi.org/10.1103/PhysRevLett.118.125501}
  (\bibinfo{year}{2017}).

\bibitem{Berman2020}
\bibinfo{author}{Berman, N.}, \bibinfo{author}{Cohen, G.} \&
  \bibinfo{author}{Fineberg, J.}
\newblock \bibinfo{journal}{\bibinfo{title}{{Dynamics and properties of the
  cohesive zone in rapid fracture and friction}}}.
\newblock {\emph{\JournalTitle{Physical Review Letters}}}
  \textbf{\bibinfo{volume}{125}}, \bibinfo{pages}{125503},
  \doiprefix\url{https://journals.aps.org/prl/abstract/10.1103/PhysRevLett.125.125503}
  (\bibinfo{year}{2020}).

\bibitem{Gvirtzman2021}
\bibinfo{author}{Gvirtzman, S.} \& \bibinfo{author}{Fineberg, J.}
\newblock \bibinfo{journal}{\bibinfo{title}{{Nucleation fronts ignite the
  interface rupture that initiates frictional motion}}}.
\newblock {\emph{\JournalTitle{Nature Physics}}} \textbf{\bibinfo{volume}{17}},
  \bibinfo{pages}{1037--1042},
  \doiprefix\url{https://doi.org/10.1038/s41567-021-01299-9}
  (\bibinfo{year}{2021}).

\bibitem{Kato2012}
\bibinfo{author}{Kato, A.} \emph{et~al.}
\newblock \bibinfo{journal}{\bibinfo{title}{{Propagation of slow slip leading
  up to the 2011 Mw 9.0 Tohoku-Oki earthquake}}}.
\newblock {\emph{\JournalTitle{Science}}} \textbf{\bibinfo{volume}{335}},
  \bibinfo{pages}{705--708},
  \doiprefix\url{https://doi.org/10.1126/science.1215141}
  (\bibinfo{year}{2012}).

\bibitem{Obara2016}
\bibinfo{author}{Obara, K.} \& \bibinfo{author}{Kato, A.}
\newblock \bibinfo{journal}{\bibinfo{title}{Connecting slow earthquakes to huge
  earthquakes}}.
\newblock {\emph{\JournalTitle{Science}}} \textbf{\bibinfo{volume}{353}},
  \bibinfo{pages}{253--257},
  \doiprefix\url{https://doi.org/10.1126/science.aaf1512}
  (\bibinfo{year}{2016}).

\bibitem{Kato2021}
\bibinfo{author}{Kato, A.} \& \bibinfo{author}{Ben-Zion, Y.}
\newblock \bibinfo{journal}{\bibinfo{title}{The generation of large
  earthquakes}}.
\newblock {\emph{\JournalTitle{Nature Reviews Earth and Environment}}}
  \textbf{\bibinfo{volume}{2}}, \bibinfo{pages}{26--39},
  \doiprefix\url{https://doi.org/10.1038/s43017-020-00108-w}
  (\bibinfo{year}{2021}).

\bibitem{Petrillo2020}
\bibinfo{author}{Petrillo, G.}, \bibinfo{author}{Lippiello, E.},
  \bibinfo{author}{Landes, F.~P.} \& \bibinfo{author}{Rosso, A.}
\newblock \bibinfo{journal}{\bibinfo{title}{{The influence of the
  brittle-ductile transition zone on aftershock and foreshock occurrence}}}.
\newblock {\emph{\JournalTitle{Nature Communications}}}
  \textbf{\bibinfo{volume}{11}}, \bibinfo{pages}{3010},
  \doiprefix\url{https://doi.org/10.1038/s41467-020-16811-7}
  (\bibinfo{year}{2020}).

\bibitem{Landau1986}
\bibinfo{author}{Landau, L.~D.}, \bibinfo{author}{Lifshitz, E.~M.},
  \bibinfo{author}{Kosevich, A.~M.} \& \bibinfo{author}{Pitaevskii, L.~P.}
\newblock \emph{\bibinfo{title}{{Theory of Elasticity}}}
  (\bibinfo{publisher}{Butterworth-Heinemann}, \bibinfo{address}{Oxford},
  \bibinfo{year}{1986}), \bibinfo{edition}{3} edn.

\bibitem{Dieterich1994}
\bibinfo{author}{Dieterich, J.~H.} \& \bibinfo{author}{Kilgore, B.~D.}
\newblock \bibinfo{journal}{\bibinfo{title}{{Direct observation of frictional
  contacts: New insights for state-dependent properties}}}.
\newblock {\emph{\JournalTitle{Pure and Applied Geophysics}}}
  \textbf{\bibinfo{volume}{143}}, \bibinfo{pages}{283--302},
  \doiprefix\url{https://doi.org/10.1007/BF00874332} (\bibinfo{year}{1994}).

\end{thebibliography}


\begin{thebibliography}{1}
\urlstyle{rm}
\expandafter\ifx\csname url\endcsname\relax
  \def\url#1{\texttt{#1}}\fi
\expandafter\ifx\csname urlprefix\endcsname\relax\def\urlprefix{URL }\fi
\expandafter\ifx\csname doiprefix\endcsname\relax\def\doiprefix{DOI: }\fi
\providecommand{\bibinfo}[2]{#2}
\providecommand{\eprint}[2][]{\url{#2}}

\bibitem{Otsuki2013}
\bibinfo{author}{Otsuki, M.} \& \bibinfo{author}{Matsukawa, H.}
\newblock \bibinfo{journal}{\bibinfo{title}{{Systematic breakdown of
  Amontons’ law of friction for an elastic object locally obeying Amontons’
  law}}}.
\newblock {\emph{\JournalTitle{Scientific Reports}}}
  \textbf{\bibinfo{volume}{3}}, \bibinfo{pages}{1586},
  \doiprefix\url{https://doi.org/10.1038/srep01586} (\bibinfo{year}{2013}).

\bibitem{Scheibert2010}
\bibinfo{author}{Scheibert, J.} \& \bibinfo{author}{Dysthe, D.~K.}
\newblock \bibinfo{journal}{\bibinfo{title}{{Role of friction-induced torque in
  stick-slip motion}}}.
\newblock {\emph{\JournalTitle{Europhysics Letters}}}
  \textbf{\bibinfo{volume}{92}}, \bibinfo{pages}{54001},
  \doiprefix\url{https://doi.org/10.1209/0295-5075/92/54001}
  (\bibinfo{year}{2010}).

\end{thebibliography}

%%%%% Acknowledgements %%%%%
\section*{Acknowledgements}

This study was supported by JSPS KAKENHI Grant Numbers JP19K03670, JP20K03792, JP21H01006, and JP22J20527. 
The numerical simulations were partially conducted on the supercomputer systems in ISSP, University of Tokyo, Japan, and in YITP, Kyoto University, Japan. 
We would like to thank Editage (www.editage.com) for English language editing.

%%%%% Author contributions statement %%%%%
\section*{Author contributions statement}

W.I. conducted the FEM simulations and analysis based on simplified models. 
All authors analyzed the results and reviewed the manuscript.

%%%%% Additional information %%%%%
\section*{Additional information}

\textbf{Supplementary information} accompanies this paper at \href{https://www.nature.com/srep}{https://www.nature.com/srep}
\vskip.5\baselineskip
\noindent \textbf{Competing interests}: The authors declare no competing interests.

\end{document}

\maketitle
\pagestyle{empty}
%%%%%%%%%%%%%%%%%% Label %%%%%%%%%%%%%%%%%%%%%%%
% \captionsetup[figure]{labelsep=period, labelfont={sf, bf}}
% \renewcommand{\figurename}{Figure}
% \renewcommand{\tablename}{Table}
% \renewcommand{\prechaptername}{}
% \renewcommand{\postchaptername}{}
% \pagenumbering{arabic} 

%%%%%%%%%%%%%%%%%%%%%%%%%%%%%%%%%%%%%%%%%%%%%%%%%%%%
%%%%%%%%%%%%%% Supplementary Video %%%%%%%%%%%%%%%%%
%%%%%%%%%%%%%%%%%%%%%%%%%%%%%%%%%%%%%%%%%%%%%%%%%%%%
\noindent{\sffamily\bfseries\Large Supplementary Video (Titles and legends)} \par
\normalsize
\vskip 1.0em

\noindent
\textsf{\textbf{Supplementary Video S1.}}~
Spatial distributions of different quantities in the frictional interface for $P_\mathrm{ext}/E = 0.012$, $L/H = 4$, and $W/H = 1$. 

\vskip 0.25em
Left panel shows the ratio $F_\mathrm{T}/F_\mathrm{N}$ against the displacement of the rigid rod $U$. 
(a)~Spatial distribution of the slip region, which is represented by the yellow area. 
(b)~Spatial distribution of the ratio $\sigma^\mathrm{(fric)}/p$. 
The white area represents the region with $p = 0$ due to the lift at the bottom. 
(c)~Spatial distribution of $p$.

\vskip 1.0em
\noindent
\textsf{\textbf{Supplementary Video S2.}}~
Spatial distributions of different quantities in the frictional interface for $P_\mathrm{ext}/E = 0.012$, $L/H = 1$, and $W/H = 4$. 

\vskip 0.25em
Left panel shows the ratio $F_\mathrm{T}/F_\mathrm{N}$ against $U$. 
(a)~Spatial distribution of the slip region, which is represented by the yellow area. 
(b)~Spatial distribution of the ratio $\sigma^\mathrm{(fric)}/p$. 
The white area represents the region with $p = 0$ due to the lift at the bottom. 
(c)~Spatial distribution of $p$.

\clearpage
\pagestyle{plain}
\setcounter{page}{1}
%%%%%%%%%%%%%%%%%%%%%%%%%%%%%%%%%%%%%%%%%%%%%%%%%%%%
%%%%%%%%%%%%%% Supplementary Note %%%%%%%%%%%%%%%%%%
%%%%%%%%%%%%%%%%%%%%%%%%%%%%%%%%%%%%%%%%%%%%%%%%%%%%
\noindent{\sffamily\bfseries\Large Supplementary Note} \par
\normalsize
\vskip 0.5em

In this supplementary note, we provide a detailed description of the analysis presented in the main text. 
Section~\ref{chap:apdx_3dfem} describes the FEM results for the spatial distribution of quantities at the frictional interface.
Details of the analytical calculation based on the simplified models are presented in Sec.~\ref{chap:apdx_1d_model}.
In Sec.~\ref{chap:apdx_compare}, we compare our results with those of a previous study~\cite{Otsuki2013}.

%%%%%%%%%%%%%%%%%%%%%%%%%%%%%%%%%%%%%%%%%%%%%%%%
%%%%%%%%%%%%%%%% Section 1 %%%%%%%%%%%%%%%%%%%%%
\vskip 1.em
\section{Spatial distributions of quantities in the frictional interface}
\label{chap:apdx_3dfem}
\vskip 0.25em
%%%%%%%%%%%%%%%%%%%%%%%%%%%%%%%%%%%%%%%%%%%%%%%%

\begin{figure}[b]
    \centering
    \includegraphics[scale=1.]{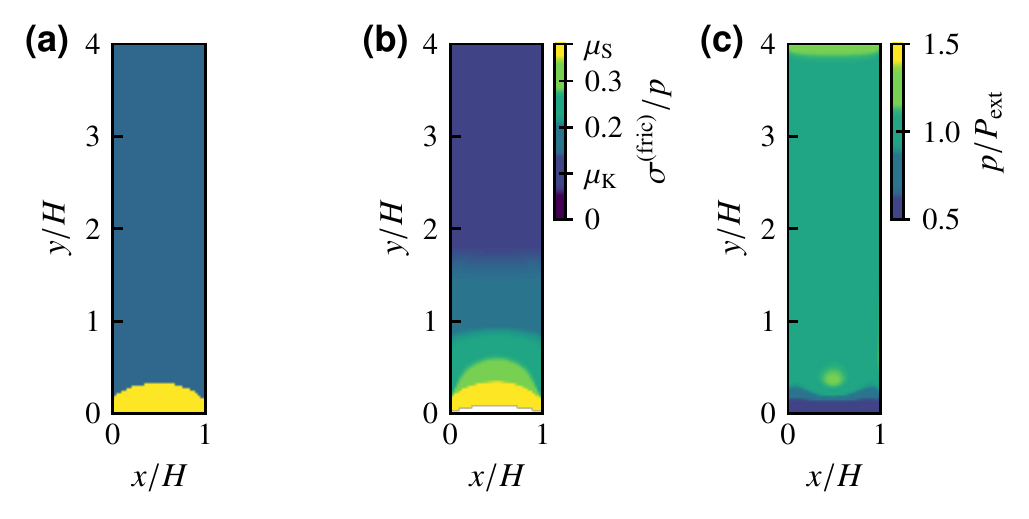}
    \caption{Spatial distributions of different quantities in the frictional interface for $P_\mathrm{ext}/E = 0.012$, $L/H = 4$, $W/H = 1$, and $U/L = 2.6 \times 10^{-2}$. 
    The rigid rod pushes the block at $(x/H,y/H)=(0.5,0)$. 
    (a)~Spatial distribution of the slip region, which is represented by the yellow area. 
    (b)~Spatial distribution of the ratio $\sigma^\mathrm{(fric)}/p$. 
    (c)~Spatial distribution of $p$. }
    \label{fig:s1_slip_sfric_p_long}
\end{figure}

\begin{figure}[!b]
    \centering
    \includegraphics[scale=1.]{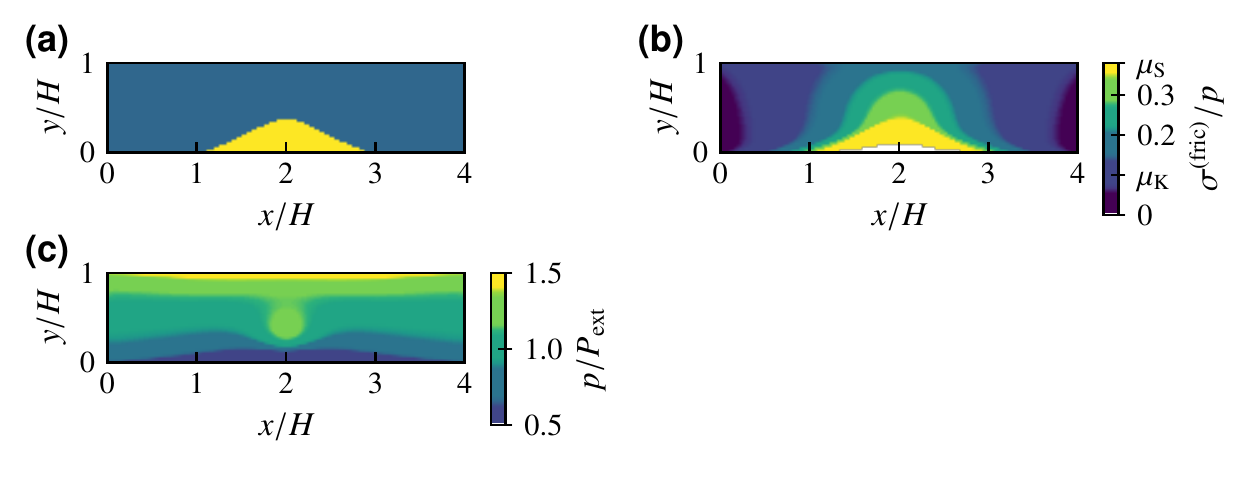}
    \caption{Spatial distributions of different quantities in the frictional interface for $P_\mathrm{ext}/E = 0.012$, $L/H = 1$, $W/H = 4$, and $U/L = 1.08 \times 10^{-1}$. 
    The rigid rod pushes the block at $(x/H,y/H)=(2,0)$. 
    (a)~Spatial distribution of the slip region, which is represented by the yellow area.
    (b)~Spatial distribution of the ratio $\sigma^\mathrm{(fric)}/p$. 
    (c)~Spatial distribution of $p$. }
    \label{fig:s2_slip_sfric_p_wide}
\end{figure}

This section presents the spatial distributions of the quantities in the frictional interface for the 3D FEM simulations, which supports the assumption used to derive the simplified models in the main text.
The spatial distributions for $L/H=4$ and $W/H=1$ are shown in Fig.~\ref{fig:s1_slip_sfric_p_long}, corresponding to Supplementary Video~S1.
Figures~\ref{fig:s1_slip_sfric_p_long}~(a) and (b) demonstrate the spatial distributions of the slip region and the ratio $\sigma^\mathrm{(fric)}/p$, respectively. 
Both distributions exhibit a slight dependency on $x$.
Therefore, in the simplified model with $L/H \gg 1$, we neglect the $x$ dependence.
The spatial distribution of pressure $p$ at the bottom is shown in Fig.~\ref{fig:s1_slip_sfric_p_long}~(c). 
The pressure $p$ is approximately equal to the pressure $P_\mathrm{ext}$ at the top surface of the entire bottom plane. 
Based on this numerical result, we assume that the bottom pressure is uniform in the simplified model for $L/H \gg 1$.

The spatial distributions in the frictional interface for $W/H=4$ with $L/H=1$ are shown in Fig.~\ref{fig:s2_slip_sfric_p_wide}, corresponding to Supplementary Video~S2.
Figures~\ref{fig:s2_slip_sfric_p_wide} (a) and (b) demonstrate the spatial distributions of the slip region and the ratio $\sigma^\mathrm{(fric)}/p$, respectively. 
The regions for precursor slip and $\sigma^\mathrm{(fric)}/p \thickapprox \mu_\mathrm{S}$ propagate from the center at $x/H=2$, which is different from those for $L/H \gg 1$ shown in Fig.~\ref{fig:s1_slip_sfric_p_long}.
The characteristic behavior is the extension of the slip region along the $x$ direction, as shown in Supplementary Video~S2. 
Therefore, in the model for $W/H \gg 1$, we neglect the dependence on $y$.
The spatial distribution of pressure $p$ at the bottom is shown in Fig.~\ref{fig:s2_slip_sfric_p_wide}~(c). 
The variation in $p$ is not strong. 
Therefore, we assume a constant pressure in the simplified model for $W/H \gg 1$.

%%%%%%%%%%%%%%%%%%%%%%%%%%%%%%%%%%%%%%%%%%%%%%%%
%%%%%%%%%%%%%%%% Section 2 %%%%%%%%%%%%%%%%%%%%%
\vskip 1.em
\section{Analysis based on the simplified models}
\label{chap:apdx_1d_model}
\vskip 0.25em
%%%%%%%%%%%%%%%%%%%%%%%%%%%%%%%%%%%%%%%%%%%%%%%%
This section presents details of the analytical calculations based on the simplified models.
We derive the results for $L/H \gg 1$ and $W/H \gg 1$ shown in Secs.~\ref{sec:apdx_1d_model_l} and \ref{sec:apdx_1d_model_w}.

%%%%%%%%%%%%%%%%%%%% 2.1 %%%%%%%%%%%%%%%%%%%%%%%
\vskip 1.0em
\subsection{Model for large \texorpdfstring{$L/H$}{L/H}}
\label{sec:apdx_1d_model_l}
%%%%%%%%%%%%%%%%%%%%%%%%%%%%%%%%%%%%%%%%%%%%%%%%

%%%%%%%%%%%%%%%%%% Adiabatic Solution %%%%%%%%%%%%%%%%%%
\subsubsection*{Quasi-static solution}
\label{subsec:quasi-static_1dl}

%%% subsec1 U = 0 %%%
The quasi-static solution $u_{y\,0}(y)$ of equations~(2) and (3) in the main text with the local friction coefficient $\mu=\mu_\mathrm{K}$ and acceleration $\ddot{u}_y(y,t)=0$ at $U=0$ is given by 
\begin{eqnarray}
    u_{y\,0}(y) = \frac{\mu_\mathrm{K} P_\mathrm{ext}}{E_1 \alpha_\mathrm{\,A} H} \left( \frac{y^2}{2} 
    - L y \right). 
    \label{eq:u0_1dl}
\end{eqnarray}
%%% subsec2 U > 0 %%%
The quasi-static solution $u^{\mathrm{(a)}}_{y}(y)$ for $U>0$ is given by
\begin{equation}
    u^{\mathrm{(a)}}_{y}(y) = \left\{
    \begin{array}{ll}
        u_{y1}(y), &~ 0 \leq y \leq l \\
        u_{y\,0}(y), &~ l < y \leq L
    \end{array}
    \right. , 
    \label{eq:ua}
\end{equation}
where $l$ is the length of the precursor slip.
The slip area $A$ is given by
\begin{equation}
    A = l W.
    \label{eq:def_l}
\end{equation}
Here, we set the local friction coefficient $\mu$ to be equal to $\mu_\mathrm{S}$ for $0 \leq y \leq l$, and the displacement and stress are continuous at $y=l$:
\begin{eqnarray}
    u_{y1}(y = l) &=& u_{y\,0}(y = l) 
    \label{eq:bc_u1_1} \\
    \left. \frac{d u_{y1}}{d y} \right|_{y=l} &=& \left. \frac{d u_{y\,0}}{d y} \right|_{y=l},
    \label{eq:bc_u1_2}
\end{eqnarray}
which gives
\begin{equation}
    u_{y1}(y) = u_{y\,0}(y) 
    + \frac{\left( \mu_\mathrm{S} - \mu_\mathrm{K} \right) P_\mathrm{ext}}{2 E_1 \alpha_\mathrm{\,A} H} 
    \left( y^2 - 2 l y + l^2 \right)
    \label{eq:u1}
\end{equation}
and
\begin{equation}
    l = \left\{ \frac{2 E_1 \alpha_\mathrm{\,A} H ~ U}{\left( \mu_\mathrm{S} - \mu_\mathrm{K} \right) P_\mathrm{ext}} 
    \right\} ^ {\frac{1}{2}}. 
    \label{eq:l_u1}
\end{equation}
From equations~\eqref{eq:def_l} and \eqref{eq:l_u1}, $A$ is obtained as 
\begin{equation}
    \frac{A}{W} = \left\{ \frac{2 E_1 \alpha_\mathrm{\,A} H ~ U}{\left( \mu_\mathrm{S} - \mu_\mathrm{K} \right) P_\mathrm{ext}} 
    \right\} ^ {\frac{1}{2}}. 
    \label{eq:A_u1}
\end{equation}

%%%%%%%%%%%%%%%%%% sec2 Stability Analysis %%%%%%%%%%%%%%%%%%
\subsubsection*{Stability analysis}
\label{subsec:stability_1dl}

We introduce the fluctuation $\delta u_{y} (y, t)$ as 
\begin{equation}
    \delta u_y(y, t) = u_y(y, t) - u^{\mathrm{(a)}}_{y}(y).
    \label{eq:def_delta_u_l}
\end{equation}
For $y > l$, we assume $\delta u_y (y, t)=0$.
Substituting equation~\eqref{eq:def_delta_u_l} into the equation of motion~(2), we obtain 
\begin{equation}
    \rho \delta \ddot{u}_y
    = E_1 \frac{\partial^2 \delta u_y}{\partial y^2} + \eta_\mathrm{\,t} \frac{\partial^2 \delta \dot{u}_y}{\partial y^2} 
    + \frac{\left( \mu_\mathrm{S} - \mu_\mathrm{K} \right) \delta \dot{u}_y ~ P_\mathrm{ext}}{v_\mathrm{c} \alpha_\mathrm{\,A} H}. 
    \label{eq:eom_sa}
\end{equation}
Because $\delta u_y = 0$ at $y=0$ and $l$, $\delta u_y(y, t)$ can be expressed as
\begin{equation}
    \delta u_y(y, t) = \sum_{m=1}{ u_{ym} e^{\lambda_m t} \sin{k_m \xi} }
    \label{eq:delta_fourier}
\end{equation}
with a positive integer $m$ and eigenvalue $\lambda_m$ of the time evolution operator, where $k_m$ and $\xi$ are defined as 
\begin{eqnarray}
    k_m &=& m \pi 
    \label{eq:km} \\
    \xi &=& \frac{y}{l},
    \label{eq:xi}
\end{eqnarray}
respectively.
Substituting equation~\eqref{eq:delta_fourier} into equation~\eqref{eq:eom_sa}, multiplying by $2\sin{k_n\xi}$ with a positive integer $n$, and integrating for $0 < y < l$, we obtain 
\begin{equation}
    \left\{ \rho \lambda_m^2 + \eta_\mathrm{\,t} \frac{k_m^2}{l^2} \lambda_m 
    + E_1 \frac{k_m^2}{l^2}
    - \frac{ \left( \mu_\mathrm{S} - \mu_\mathrm{K} \right) P_\mathrm{ext} }
    {v_\mathrm{c} \alpha_\mathrm{\,A} H} \lambda_m  \right\} 
    u_n e^{\lambda_m t} = 0. 
    \label{eq:lambda_ver3_1}
\end{equation}
From equations~\eqref{eq:def_l} and \eqref{eq:lambda_ver3_1}, we obtain
\begin{eqnarray}
    \rho L^2 \lambda_m^2 + k_m^2 \eta_\mathrm{\,t} \left( \frac{A}{A_0} \right)^{-2} \lambda_m 
    + k_m^2 E_1 \left( \frac{A}{A_0} \right)^{-2}
    - \frac{ \left( \mu_\mathrm{S} - \mu_\mathrm{K} \right) P_\mathrm{ext} L^2 }
    {v_\mathrm{c} \alpha_\mathrm{\,A} H} \lambda_m = 0. 
    \label{eq:lambda_ver3}
\end{eqnarray}
The second term on the left-hand side of equation~\eqref{eq:lambda_ver3} corresponds to the viscosity, while the fourth term on the same side represents the velocity-weakening friction. 
For $A/A_0 \ll 1$, we can neglect the fourth term and $\lambda_m$ satisfies $\Re \lambda_m < 0$.
For a larger $A/A_0$, the second and third terms can be neglected, and one of the solutions satisfies $\Im \lambda_m = 0 ~~\mathrm{and}~~ \Re \lambda_m > 0$.

The fluctuation $\delta u_y(y,t)$ becomes unstable when $\Re \lambda_m > 0$; however, the oscillatory instability with $\Im \lambda_m \neq 0$ is suppressed by the static friction force because the oscillatory motion is accompanied by backward motion. 
Bulk sliding occurs when the eigenvalue satisfies
\begin{eqnarray}
    \Im \lambda_m = 0 ~~\mathrm{and}~~ \Re \lambda_m > 0
    \label{eq:lambda_c1}
\end{eqnarray}
with $m=1$. 
We obtain equation~(4) in the main text for the critical area $A_\mathrm{c}$ from this condition using equation~\eqref{eq:lambda_ver3}, which gives
\begin{eqnarray}
    \frac{A_\mathrm{c}}{A_0} 
    = \dfrac{\pi \eta_\mathrm{\,t}}{ L \left( - \sqrt{ \rho E_1 } + \sqrt{ \rho E_1 +  
    \dfrac{ \left( \mu_\mathrm{S} - \mu_\mathrm{K} \right) P_\mathrm{ext} \eta_\mathrm{\,t} }{v_\mathrm{c} \alpha_\mathrm{\,A} H } } \right) }.
    \label{eq:lc_p_1dl_ver2}
\end{eqnarray}

%%%%%%%%%%%%%%%%%% sec3 Friction Coefficient %%%%%%%%%%%%%%%%%%
\subsubsection*{Macroscopic friction coefficient}
\label{subsec:mu_1dl}

The loading force $F_\mathrm{N}$ is given by
\begin{equation}
    F_\mathrm{N} = P_\mathrm{ext} \,A_0. 
    \label{eq:FN}
\end{equation}
The driving force $F_\mathrm{T}$ is balanced with the total friction force, except for the duration of bulk sliding.  
It should be noted that the precursor motion is quasi-static.
$F_\mathrm{T}$ is given by the integral of the local friction force as 
\begin{equation}
    F_\mathrm{T} = W \int_0^{L} { \mu P_\mathrm{ext} dy}.
    \label{eq:FT_ver1}
\end{equation}
Using equations~(2), (3), \eqref{eq:u0_1dl}, \eqref{eq:def_l}, \eqref{eq:ua}, \eqref{eq:bc_u1_2}, and \eqref{eq:u1}, we obtain 
\begin{eqnarray}
    F_\mathrm{T} &=& \alpha_\mathrm{\,A} H W \int_0^{L} { \frac{d \sigma_{yy}(y)}{d y} dy} 
    \nonumber \\
    &=& E_1 \alpha_\mathrm{\,A} H W \left\{ \int_0^{l} { \frac{d^2 u_{y1}}{d y^2} dy} 
    + \int_l^{L} { \frac{d^2 u_{y0}}{d y^2} dy} \right\}
    \nonumber \\
    &=& - E_1 \alpha_\mathrm{\,A} H W \left. \frac{d u_{y1}}{d y} \right|_{y=0}
    \nonumber \\
    &=& P_\mathrm{ext} \,A_0 \left\{ \mu_\mathrm{K} + \left( \mu_\mathrm{S} - \mu_\mathrm{K} \right) \frac{A}{A_0} \right\}. 
    \label{eq:FT_ver2}
\end{eqnarray}
Because the macroscopic static friction coefficient $\mu_\mathrm{M}$ is the ratio of the driving force to the loading force, $F_\mathrm{T}/F_\mathrm{N}$, at $A=A_\mathrm{c}$, equation~(6) in the main text is obtained from equations~\eqref{eq:FN} and \eqref{eq:FT_ver2}.

%%%%%%%%%%%%%%%%%%%% 2.2 %%%%%%%%%%%%%%%%%%%%%%%
\vskip 1.0em
\subsection{Model for large \texorpdfstring{$W/H$}{W/H}}
\label{sec:apdx_1d_model_w}
%%%%%%%%%%%%%%%%%%%%%%%%%%%%%%%%%%%%%%%%%%%%%%%%

%%%%%%%%%%%%%%%%%% Adiabatic Solution %%%%%%%%%%%%%%%%%%
\subsubsection*{Quasi-static solution}
\label{subsec:quasi-static_1dw}

%%% subsec1 U = 0 %%%
The quasi-static solution $u_{y\,0}(x)$ of equations~(9) and (10) in the main text at $U=0$ is obtained as
\begin{eqnarray}
    u_{y\,0}(x) = \frac{\mu_\mathrm{K} P_\mathrm{ext}}{2 E_2 \alpha_\mathrm{\,B} H}\left( x^2 - W x \right)
    \label{eq:u0_1dw}
\end{eqnarray}
with the boundary conditions, local friction coefficient $\mu=\mu_\mathrm{K}$, and acceleration $\ddot{u}_y(x,t)=0$.
%%% subsec2 U > 0 %%%
For $U>0$, the quasi-static solution $u^{\mathrm{(a)}}_{y}(x)$ is given by 
\begin{equation}
    u^{\mathrm{(a)}}_{y}(x) = \left\{
    \begin{array}{ll}
        u_{y1} (x), &~ 0 \leq x \leq \frac{w}{2} \\
        u_{y\,0} (x), &~ \frac{w}{2} < x \leq \frac{W}{2} 
    \end{array}
    \right.
    \label{eq:ua_1dw}
\end{equation}
with the width $w$ of the precursor slip related to $A$ as
\begin{equation}
    A = w L. 
    \label{eq:def_w}
\end{equation}
Setting $\mu$ equal to $\mu_\mathrm{S}$ for $0 \leq x \leq w/2$, the displacement and stress are continuous at $x=w/2$:
\begin{eqnarray}
    u_{y1} \left( x = \tfrac{w}{2} \right) &=& u_{y\,0} \left( x = \tfrac{w}{2} \right) 
    \label{eq:bc_u1_l} \\
    \left. \frac{d u_{y1}}{d x} \right|_{x=\frac{w}{2}} &=& 
    \left. \frac{d u_{y\,0}}{d x} \right|_{x=\frac{w}{2}},
    \label{eq:bc_e1_l}
\end{eqnarray}
where we obtain $u_{y1}(x)$ as 
\begin{equation}
    u_{y1}(x) 
    = u_{y\,0}(x) + \frac{ (\mu_\mathrm{S} - \mu_\mathrm{K}) P_\mathrm{ext} }
    { 2 E_2 \alpha_\mathrm{\,B} H } \left( x^2 - w x + \frac{w^2}{4} \right) 
    \label{eq:u1_1dw}
\end{equation}
with 
\begin{equation}
    w = 2 \left\{ \frac{ 2 E_2 \alpha_\mathrm{\,B} H U }{ (\mu_\mathrm{S} - \mu_\mathrm{K}) P_\mathrm{ext} } \right\}^{\frac{1}{2}}. 
    \label{eq:l_1dw}
\end{equation}
From equation~\eqref{eq:def_w}, $A$ can be expressed as 
\begin{equation}
    \frac{A}{L} = 2 \left\{ \frac{ 2 E_2 \alpha_\mathrm{\,B} H U }{ (\mu_\mathrm{S} - \mu_\mathrm{K}) P_\mathrm{ext} } \right\}^{\frac{1}{2}}. 
    \label{eq:A_1dw}
\end{equation}

%%%%%%%%%%%%%%%%%% sec2 Stability Analysis %%%%%%%%%%%%%%%%%%
\subsubsection*{Stability analysis}
\label{subsec:stability_1dw}

We introduce the fluctuation $\delta u_y(x, t)$ as 
\begin{equation}
    \delta u_y(x, t) = u_y(x, t) - u^{\mathrm{(a)}}_{y}(x).
    \label{eq:def_delta_u_1dw}
\end{equation}
Substituting equation~\eqref{eq:def_delta_u_1dw} into equation~(9), we obtain 
\begin{equation}
    \rho \delta \ddot{u}_y
    = E_2 \frac{\partial^2 \delta u_y}{\partial x^2} + \frac{\eta_1}{2} \frac{\partial^2 \delta \dot{u}_y}{\partial x^2} 
    + \frac{ \left( \mu_\mathrm{S} - \mu_\mathrm{K} \right) P_\mathrm{ext} \delta \dot{u}_y }
    {v_\mathrm{c} \alpha_\mathrm{\,B} H}. 
    \label{eq:eom_sa_1dw}
\end{equation}
Assuming $\delta u_y (x, t)=0$ for $x > w/2$ and $x=0$, $\delta u_y(x, t)$ is expressed as 
\begin{equation}
    \delta u_y(x, t) = \sum_{m=1}{ u_{ym} e^{\lambda_m t} \sin{k_m \xi} }, 
    \label{eq:delta_fourier_1dw}
\end{equation}
where $k_m$ and $\xi$ are given by 
\begin{eqnarray}
    k_m &=& m \pi 
    \label{eq:apdx_km_1dw} \\
    \xi &=& \frac{2 x}{w}. 
    \label{eq:xi_1dw}
\end{eqnarray}
Here, $m$ is a positive integer and $\lambda_m$ is the eigenvalue. 
Substituting equation~\eqref{eq:delta_fourier_1dw} into equation~\eqref{eq:eom_sa_1dw}, multiplying by $2\sin{k_n\xi}$, and integrating for $0 < x < w/2$, we obtain
\begin{equation}
    \left\{ \rho \lambda_m^2 + 2 \eta_1 \frac{k_m^2}{w^2} \lambda_m 
    + 4 E_2 \frac{k_m^2}{w^2} 
    - \frac{ \left( \mu_\mathrm{S} - \mu_\mathrm{K} \right) P_\mathrm{ext} }
    { v_\mathrm{c} \alpha_\mathrm{\,B} H } \lambda_m \right\} 
    u_n e^{\lambda_m t} = 0. 
    \label{eq:lambda_1dw_ver2_1}
\end{equation}
From equations~\eqref{eq:def_w} and \eqref{eq:lambda_1dw_ver2_1}, we obtain
\begin{equation}
    \rho W^2 \lambda_m^2 + 2 k_m^2 \eta_1 \left( \frac{A}{A_0} \right)^{-2} \lambda_m 
    + 4 k_m^2 E_2 \left( \frac{A}{A_0} \right)^{-2} 
    - \frac{ \left( \mu_\mathrm{S} - \mu_\mathrm{K} \right) P_\mathrm{ext} W^2}
    { v_\mathrm{c} \alpha_\mathrm{\,B} H } \lambda_m = 0. 
    \label{eq:lambda_1dw_ver2}
\end{equation}
The second term on the left-hand side of equation~\eqref{eq:lambda_1dw_ver2} represents the viscosity, while the fourth term on the same side originates from the velocity-weakening friction.

Similar to the case for $L/H \gg 1$, bulk sliding occurs when the eigenvalue satisfies equation~\eqref{eq:lambda_c1} with increasing $A/A_0$ owing to the instability caused by the velocity-weakening friction. 
From this condition, we obtain equation~(11) in the main text for the critical area $A_\mathrm{c}$, which gives
\begin{eqnarray}
    \frac{A_\mathrm{c}}{A_0} = \frac{ \pi \eta_1 }{ W \left( - \sqrt{ \rho E_2 } + \sqrt{ \rho E_2 + 
    \dfrac{ (\mu_\mathrm{S} - \mu_\mathrm{K}) P_\mathrm{ext} \eta_1 }{ 2 v_\mathrm{c} \alpha_\mathrm{\,B} H } } \right) }. 
    \label{eq:lc_p_1dw_ver2}
\end{eqnarray}

%%%%%%%%%%%%%%%%%% sec3 Friction Coefficient %%%%%%%%%%%%%%%%%%
\subsubsection*{Macroscopic friction coefficient}
\label{subsec:mu_1dw}

The loading force $F_\mathrm{N}$ is given by equation~\eqref{eq:FN}.
The driving force $F_\mathrm{T}$ is balanced with the integral of the local friction force as 
\begin{equation}
    F_\mathrm{T} = L \int_{-\frac{W}{2}}^{\frac{W}{2}} { \mu P_\mathrm{ext} dx} 
    = 2 L \int_0^{\frac{W}{2}} { \mu P_\mathrm{ext} dx}.
    \label{eq:FT_1dw_ver,}
\end{equation}
Using equations~(9), (10), \eqref{eq:u0_1dw}, \eqref{eq:def_w}, \eqref{eq:ua_1dw}, \eqref{eq:bc_e1_l}, and \eqref{eq:u1_1dw}, we obtain 
\begin{eqnarray}
    F_\mathrm{T} &=& 2 \alpha_\mathrm{\,B} H L \int_0^{\frac{W}{2}} { \frac{d \sigma_{xy}(x)}{dx} dx} 
    \nonumber \\
    &=& 2 E_2 \alpha_\mathrm{\,B} H L \left\{ \int_0^{\frac{w}{2}} { \frac{d^2 u_{y1}}{dx^2} dx} 
    + \int_{\frac{w}{2}}^{\frac{W}{2}} { \frac{d^2 u_{y0}}{d x^2} dx} \right\}
    \nonumber \\
    &=& - 2 E_2 \alpha_\mathrm{\,B} H L \left. \frac{d u_{y1}}{d x} \right|_{x=0}
    \nonumber \\
    &=& P_\mathrm{ext} \,A_0 
    \left\{ \mu_\mathrm{K} + \left( \mu_\mathrm{S} - \mu_\mathrm{K} \right) \frac{A}{A_0} \right\}. 
    \label{eq:FT_1dw_ver2}
\end{eqnarray}
Because the macroscopic static friction coefficient $\mu_\mathrm{M}$ is the ratio of the driving force to the loading force, $F_\mathrm{T}/F_\mathrm{N}$, at $A=A_\mathrm{c}$, we obtain equation~(6) in the main text from equations~\eqref{eq:FN} and \eqref{eq:FT_1dw_ver2}.

%%%%%%%%%%%%%%%%%%%%%%%%%%%%%%%%%%%%%%%%%%%%%%%%
%%%%%%%%%%%%%%%% Section 3 %%%%%%%%%%%%%%%%%%%%%
\vskip 1.em
\section{Comparison with the analytical result from a previous study}
\label{chap:apdx_compare}
\vskip 0.25em
%%%%%%%%%%%%%%%%%%%%%%%%%%%%%%%%%%%%%%%%%%%%%%%%

In this section, we compare our results with those of ref.~\citenum{Otsuki2013}. 
A previous study discussed the dependence of the precursor slip and friction coefficient on system size with a constant aspect ratio $L/H=2$. 
In the analysis, the authors considered the nonuniformity of the bottom pressure as the origin of the precursor slip and demonstrated the breakdown of Amontons' law. 
The bottom pressure was shown to increase along the driving direction owing to the torque effect~\cite{Otsuki2013, Scheibert2010} and was approximated by $p(y) = 2 P_\mathrm{ext} \,y/L$. 
The equation of motion is given by
\begin{equation}
    \rho \ddot{u}_y(y,t) = \frac{\partial \sigma_{yy}(y,t)}{\partial y}
    + \frac{\sigma_{zy}(y,t)-\mu(\dot{u}_y) p(y)}{\alpha H}, 
    \label{eq:eom_2013}
\end{equation}
where the stresses $\sigma_{yy}$ and $\sigma_{zy}$ are expressed as 
\begin{eqnarray}
    \sigma_{yy}&=&E_1 \frac{\partial u_y}{\partial y} + \eta_\mathrm{\,t} \frac{\partial \dot{u}_y}{\partial y}
    \label{eq:sigma_yy_2013} \\
    \sigma_{zy}&=&E_2 \frac{U-u_y}{H/2} + \frac{\eta_1}{2}\frac{V_\mathrm{rod} - \dot{u}_y}{H/2}
    \label{eq:sigma_zy_2013}
\end{eqnarray}
with the fitting parameter $\alpha$. 
The boundary conditions are as 
\begin{equation}
    \left. \frac{\partial u_y}{\partial y} \right|_{y=0} = \left. \frac{\partial u_y}{\partial y} \right|_{y=L} = 0. 
    \label{eq:bc_2013}
\end{equation}
Using the same procedure as that in Sec.~\ref{chap:apdx_1d_model}, the equation for the critical area $A_\mathrm{c}$ is obtained as 
\begin{eqnarray}
    && \frac{\pi^2 \eta_\mathrm{\,t}}{4} \left( \frac{A_\mathrm{c}}{A_0} \right)^{-2}
    + \frac{\eta_1}{\alpha} \left( \frac{L}{H} \right)^{2}
    + 2 L \sqrt{ \rho \left\{ \frac{\pi^2 E_1}{4} \left( \frac{A_\mathrm{c}}{A_0} \right)^{-2} + \frac{2 E_2}{\alpha} \left( \frac{L}{H} \right)^2 \right\} } \nonumber\\
    &=& \frac{(\pi^2 - 4)}{\pi^2}
    \frac{(\mu_\mathrm{S} - \mu_\mathrm{K}) P_\mathrm{ext} L^2}{\alpha v_\mathrm{c} H}
    \frac{A_\mathrm{c}}{A_0}. 
    \label{eq:Ac_equal_2013}
\end{eqnarray}
For $A_\mathrm{c}/A_0 \ll 1$, $A_\mathrm{c}/A_0$ is expressed as 
\begin{equation}
    \frac{A_\mathrm{c}}{A_0} \simeq \frac{\pi^2}{\pi^2 - 4}
    \left( \frac{\mu_\mathrm{S} - \mu_\mathrm{K}}{\alpha} \right)^{-\frac{1}{3}}
    \left( \frac{P_\mathrm{ext} H}{\eta_\mathrm{\,t} v_\mathrm{c}} \right)^{-\frac{1}{3}}
    \left( \frac{L}{H} \right)^{-\frac{2}{3}},
    \label{eq:Ac_A0_2013}
\end{equation}
where the exponents for $P_\mathrm{ext}$ and $L/H$ differ from those in equation~(5) in the main text. 
The difference in the exponents results from the different assumptions for the bottom pressure adopted in both studies, i.e., uniform pressure in the present study and non-uniform pressure in the previous study. 
Note that the macroscopic friction coefficient $\mu_\mathrm{M}$ in this model is also given by equation~(6) in the main text.

%%%
\begin{figure}[!tb]
    \centering
    \includegraphics[scale=1.]{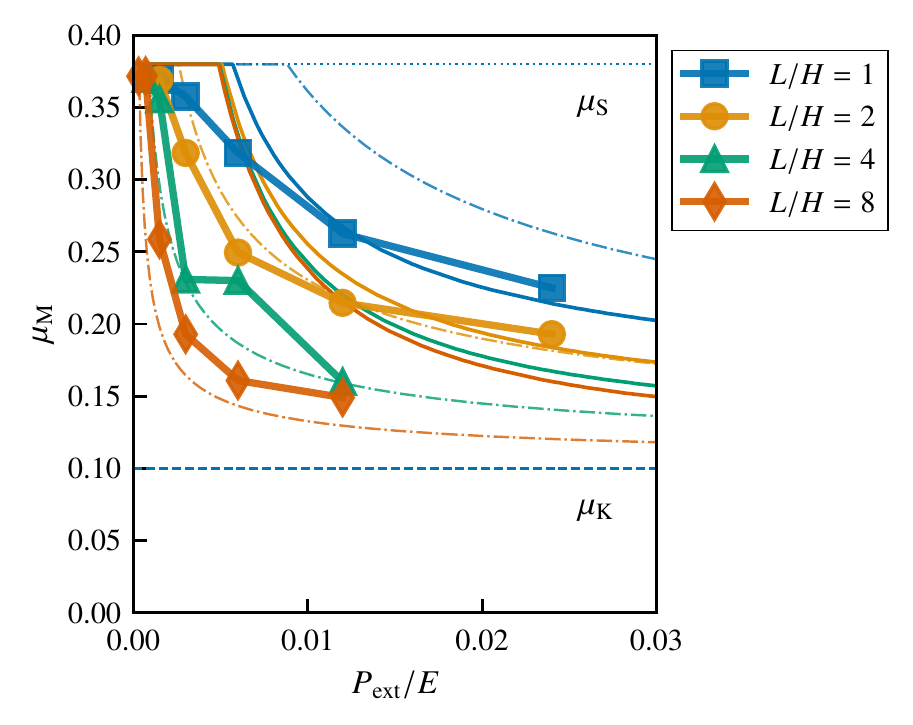}
    \caption{Dependence of $\mu_\mathrm{M}$ on $P_\mathrm{ext}$ for various $L/H$ values with $W/H = 1$. 
    The thin solid lines represent the analytical results in ref.~\citenum{Otsuki2013} given by equations~\eqref{eq:Ac_equal_2013} and (6) with the fitting parameter $\alpha = 0.05$. 
    The thin dash-dotted lines represent the analytical results in the present study given by equations~\eqref{eq:lc_p_1dl_ver2} and (6) with the fitting parameter $\alpha_\mathrm{A} = 0.2$.
    The dotted and dashed lines represent $\mu_\mathrm{S}$ and $\mu_\mathrm{K}$, respectively.}
    \label{fig:s3_mu_pext_long_2013}
\end{figure}
%%%
In Fig.~\ref{fig:s3_mu_pext_long_2013}, we show $\mu_\mathrm{M}$ against $P_\mathrm{ext}$ for various lengths, $L/H$ with $W/H = 1$, obtained from the FEM simulations in the present study. 
The analytical results in ref.~\citenum{Otsuki2013} given by equations~\eqref{eq:Ac_equal_2013} and (6), and in the present study given by equations~\eqref{eq:lc_p_1dl_ver2} and (6) are also shown. 
The fitting parameter $\alpha$ is determined such that the analytical result matches the FEM results for $L/H=2$ and $P_\mathrm{ext}/E>0.01$. 
Figure~\ref{fig:s3_mu_pext_long_2013} shows that the analytical results in the previous study better reproduce the results of the FEM analysis for $L/H \leq 2$. 
This is because the nonuniformity of the bottom pressure is significant for $L/H \leq 2$.
However, the analytical results deviate from the FEM analysis for $L/H \geq 4$, where the bottom pressure is almost uniform, as shown in Fig.~\ref{fig:s1_slip_sfric_p_long}~(c) and Supplementary Video~S1~(c).
Instead, the analytical results of the present study agree with the FEM analysis for $L/H \geq 4$, as shown in Fig.~3 and Fig.~\ref{fig:s3_mu_pext_long_2013}.

%%%%%%%%%%%%%%%% References %%%%%%%%%%%%%%%%%%%%
\vskip 1.0em
% \printbibliography[title=References]
% \addcontentsline{toc}{chapter}{References}
\bibliography{bib_imo2022_supple}